\documentclass[prd,superscriptaddress,altaffilletter,nofootinbib]{revtex4}
\usepackage[dvips]{graphicx}
\usepackage{amsmath}
\usepackage{graphicx,epsfig}
\newcommand{\be}{\begin{equation}}
\newcommand{\ee}{\end{equation}}
\newcommand{\bea}{\begin{eqnarray}}
\newcommand{\eea}{\end{eqnarray}}
\newcommand{\der}{\partial}

\begin{document}

%%%%%%%%%%%%%%%%%%%%%%%%%%%%%%%%%%%%%%%%%%%%%%%%%%%%%%%%%%%%%%%%%%%%%%%%%%%%%%%%%%%%%%%%%%%%%%%%%%%%%%%%%%%%%%%%%%
%%%%%%%%%%%%%%%%%%%%%%%%%%%%%%%%%%%%%%%%%%%%%%%%%%%%%%%%%%%%%%%%%%%%%%%%%%%%%%%%%%%%%%%%%%%%%%%%%%%%%%%%%%%%%%%%%%

\title{Avoiding the cosmological constant issue in a class of phenomenologically viable $F(R,{\cal G})$ theories}

%%%%%%%%%%%%%%%%%%%%%%%%%%%%%%%%%%%%%%%%%%%%%%%%%%%%%%%%%%%%%%%%%%%%%%%%%%%%%%%%%%%%%%%%%%%%%%%%%%%%%%%%%%%%%%%%%

\author{Israel Quiros}\email{iquiros@fisica.ugto.mx}
\affiliation{Dpto. Ingenier\'ia Civil, Divisi\'on de Ingenier\'ia, Universidad de Guanajuato, Gto., CP 36000, M\'exico.}

%%%%%%%%%%%%%%%%%%%%%%%%%%%%%%%%%%%%%%%%%%%%%%%%%%%%%%%%%%%%%%%%%%%%%%%%%%%%%%%%%%%%%%%%%%%%%%%%%%%%%%%%%%%%%%%%%%%%%%%%

\begin{abstract}
In this paper we investigate a class of phenomenologically viable $F(R,{\cal G})$ theories that are able to avoid the cosmological constant issue. While the absence of ghosts and other kinds of instability issues is of prime importance, other reasonable requirements such as vanishing effective (low curvature) cosmological constant, including the flat space as a stable vacuum solution, are also imposed on the viable models. These are free of the cosmological constant problem thanks to the following outstanding feature: the de Sitter space is an attractor of the asymptotic cosmological dynamics, with the resulting constant Hubble rate being unrelated both to the energy density of vacuum and to the low-curvature effective cosmological constant. 
\end{abstract}

%\date{\today}

\maketitle

%%%%%%%%%%%%%%%%%%%%%%%%%%%%%%%%%%%%%%%%%%%

\section{Introduction}\label{sect-intro}

%%%%%%%%%%%%%%%%%%%%%%%%%%%%%%%%%%%%%%%%%%%

The cosmological constant problem (CCP) \cite{ccp-weinberg, ccp-peebles, ccp-padma, ccp-zlatev, ccp-carroll, ccp-sahni, ccp-staro, ccp-dreitl} is one of the current unsolved puzzles in fundamental physics. In the most widespread version of the issue, the challenge is to explain the origin of the large discrepancy between the theoretically predicted value of the energy density of vacuum $\rho^\text{theor}_\text{vac}\sim 10^{72}$ GeV$^4$ and the observed value $\rho^\text{obs}_\text{vac}\sim 10^{-48}$ GeV$^4$. 

%------------------------------aim---------------------------------

In this paper we shall not search for a solution to the CCP, that has shown to be a very complex issue with strong roots in the particle's physics sector of the field theory. Instead, we shall look for theoretically consistent modifications of general relativity (GR) that are able to avoid the issue. The absence of the problem can be an alternative explanation to the unsolved puzzle. In order to ensure this goal, the following necessary and sufficient conditions should be satisfied.

\begin{itemize}

\item{\it Necessary condition}: de Sitter space with constant Hubble rate $H=H_0$, where $H_0$ is its present value, should be an attractor of the asymptotic dynamics of the related Friedmann-Robertson-Walker (FRW) cosmological model. 

\item{\it Sufficient condition}: $H_0$ should be unrelated both to the quantum vacuum energy density and to the low-curvature effective cosmological constant (if different from the energy density of vacuum). 

\end{itemize} The necessary condition ensures that, no matter which modification of GR one is dealing with, at present it should be indistinguishable from the $\Lambda$CDM cosmological model, and that this de Sitter stage is a natural outcome of the cosmological evolution, quite independent of the chosen initial conditions. The sufficient condition assures that the present (observed) value of the Hubble rate $H_0\approx 10^{-10}h$ yr$^{-1}$, where $h$ is a dimensionless parameter in the range $0.62\lesssim h\lesssim 0.82$ or, in Planck units: $H_0\sim 10^{-42}$ GeV, has nothing to do neither with the vacuum energy density $\rho_\text{vac}\sim\rho_\text{Pl}\sim 10^{72}$ GeV$^4$ $\rightarrow$ $H_\text{vac}\sim M_\text{Pl}\sim 10^{18}$ GeV, nor with the effective (low-curvature) cosmological constant $\Lambda_\text{eff}$ (assuming that these are not coincident), which may be assumed to be vanishing if flat space is to be a solution of the equations of motion (EOM).

The phenomenologically viable models should satisfy additional reasonable consistency requirements:

\begin{enumerate}

\item The theoretical framework should be free of ghosts and other harmful instabilities.

\item For sufficiently small curvature the theory should be indistinguishable from GR with an effective -- presumably very small or even vanishing -- cosmological constant.

\item Although in curved space the energy density of the quantum vacuum must be non-vanishing, in flat space it should be zero due to some (yet undiscovered) symmetry. Hence, flat space should be a stable solution of the vacuum EOM.

\end{enumerate} The first consistency requirement above is an unavoidable theoretical criterion that any viable model of actual physical processes should respect. Ghosts that arise in modified gravity theories describe physical excitations that are drawn as external lines in Feynman diagrams \cite{clifton-phys-rept-2012}. The existence of physical ghost leads, eventually, to either the existence of negative norm states or to negative energy eigenstates. Hence, one is faced either with problems for the formulation of a consistent quantum theory or with catastrophic instabilities when the ghost couples to conventional matter fields. The second requirement impacts directly the phenomenological viability of the theoretical framework. It reflects our belief, deeply rooted in the existing amount of experimental evidence, that any modification of gravity in the Solar system, at leading order, must be very close to GR. In this regard, the third requirement is a consequence of our understanding that weak gravity may be viewed as a small deformation of Minkowski space or, in other words, that, for an isolated source of gravity, the space is asymptotically flat as in GR. One example of a theory that certainly does not satisfy one the conditions stated above: the sufficient condition for avoidance of the cosmological constant issue, is precisely general relativity.

%---------4th order curvature contribution----------------------------

Physically motivated modifications of GR can be based on the inclusion of higher-order curvature operators. Indeed, such a generalization might be considered desirable as it will cause the graviton propagator to fall off more quickly in the UV, thereby improving the renormalisability properties \cite{clifton-phys-rept-2012}. Modifying gravity in this way, however, also has a number of drawbacks. In particular, it can introduce instabilities into the theory, such as ghost-like degrees of freedom. One physically motivated example of the inclusion of higher-order curvature invariants (and their admissible combinations) is string theory. The string effective action contains an infinite, well organized and ghost-free series of higher curvature corrections to the leading Einstein gravity \cite{veltman-74, goroff-86, gross-86, deser-86, donoghue-1, donoghue-2, alvarez-gaume}. One way to incorporate the quadratic contributions to the effective action while keeping the theory ghost-free is to consider the Gauss-Bonnet invariant \cite{nojiri-1, nojiri-2, nojiri-ghfree-grav, calcagni, clifton}: ${\cal G}=R^2-4R_{\mu\nu}R^{\mu\nu}+R_{\mu\nu\sigma\lambda}R^{\mu\nu\sigma\lambda}$. However, since this term in the action amounts to a total derivative that does not affect the equations of motion, the only way in which it may affect the local dynamics of fields in 4 dimensions, is to dynamically couple it through, for instance, a scalar field \cite{nojiri-1}, or to consider general functions $F({\cal G})$ in the action \cite{nojiri-2}.

In this paper we shall investigate a class of $F(R,{\cal G})$ models where the curvature invariants enter in the following combination: $\alpha R+\beta{\cal G}$ $\Rightarrow$ $F(R,{\cal G})=F(R+c{\cal G})$, where $c=\beta/\alpha$ is a constant. We shall study, in particular, a Born-Infeld (BI) inspired class of models where the mentioned combination is within a square root \cite{bi-1934, bi-fiorini, bi-rev, comelli, prd-2010}. This particular class of Lagrangian obeys the necessary and sufficient conditions for avoidance of the CCP as well as the additional reasonable consistency requirements stated above so that, the cosmological constant issue is effectively avoided. 

The paper has been organized in the following way. In the next section \ref{sect-setup} we discuss the fundamentals of $F(R,{\cal G})=F(R+c{\cal G})$ theories, including the equations of motion and the small curvature limit. Then, in section \ref{sect-ghost-free}, we check the $F(R,{\cal G})$ theories in general to the absence of ghosts. The absence of ghosts due to the anisotropy of space is linked with the specific form of theories $F(R,{\cal G})=F(R+c{\cal G})$. In section \ref{sect-models} we concentrate in this particular class, focusing to the BI inspired model that satisfies the necessary and sufficient conditions for the avoidance of the cosmological constant issue, as well as the additional requirements for phenomenological viability. The asymptotic cosmological dynamics of the class of BI inspired models is investigated in section \ref{sect-bigb-2d} for the particular case when the vacuum has vanishing energy density, while the asymptotic dynamics of the general case when the energy density of vacuum is non-vanishing, is discussed in section \ref{sect-bigb-3d}. In both cases the late time de Sitter attractor is identified and fully characterized. The way in which the cosmological constant problem is avoided in the chosen class of models, is discussed in section \ref{sect-ccp}. Other interesting aspects of the model are discussed in section \ref{sect-discuss} while brief conclusions are given in section \ref{sect-conclu}.

%%%%%%%%%%%%%%%%%%%%%%%%%%%%%%%%%%%%%%%%%%%%%%%%%%%%%%%%%%%%%%%%%%%%%%%%%

\section{$F(R,{\cal G})$ modifications of gravity}\label{sect-setup}

%%%%%%%%%%%%%%%%%%%%%%%%%%%%%%%%%%%%%%%%%%%%%%%%%%%%%%%%%%%%%%%%%%%%%%%%%

Here we shall focus in $F(R,{\cal G})$ theories of the kind $F(Lovelock)$ gravity \cite{lovelock, bueno-lov}, i. e., 

\bea F(R,{\cal G})=F(\alpha R+\beta{\cal G}),\label{frg-gen}\eea where $\alpha$ and $\beta$ are parameters with mass dimensions $M^{-2}$ and $M^{-4}$, respectively. Hence, the following relationships take place:

\bea F_{\cal G}=\frac{\beta}{\alpha}F_R,\;F_{{\cal G}R}=F_{R{\cal G}},\;F_{RR}=\frac{\alpha}{\beta}F_{{\cal G}R},\;F_{\cal GG}=\frac{\beta}{\alpha}F_{R{\cal G}},\label{rel}\eea where $F_R\equiv\der F/\der R$, $F_{{\cal G}R}\equiv\der^2F/\der{\cal G}\der R$, etc. 

We consider an action of the form:

\bea S=\frac{1}{2\kappa^2}\int d^4x\sqrt{-g}\left[F(R,{\cal G})+2\epsilon\kappa^2\mu^4\right],\label{frg-action}\eea where $\mu$ and $\kappa$ are free parameters with the dimension of mass and inverse mass, respectively, while $\epsilon=\pm 1$. The EOM that are derived from the above action -- plus a matter piece action -- read \cite{defelice-prd-2010, odintsov-npb-2019}:

\bea &&G_{\mu\nu}+\Sigma^\text{curv}_{\mu\nu}=\kappa^2_\text{eff}\left[T^{(m)}_{\mu\nu}+\epsilon\mu^4g_{\mu\nu}\right],\label{feq}\eea where we have defined the effective gravitational coupling

\bea \kappa^2_\text{eff}\equiv\frac{\kappa^2}{F_R},\label{k2-eff}\eea while

\bea &&\Sigma^\text{curv}_{\mu\nu}=-\frac{(1+\frac{2\beta}{\alpha}R)}{F_R}(\nabla_\mu\nabla_\nu-g_{\mu\nu}\nabla^2)F_R+\frac{1}{2}\left(R+\frac{\beta}{\alpha}{\cal G}-\frac{F}{F_R}\right)g_{\mu\nu}\nonumber\\
&&\;\;\;\;\;\;\;\;\;\;\;+\frac{4\beta}{\alpha F_R}\left(R_{\lambda\mu}\nabla^\lambda\nabla_\nu+R_{\lambda\nu}\nabla^\lambda\nabla_\mu-R_{\mu\nu}\nabla^2\right)F_R+\frac{4\beta}{\alpha}\,\left(R_{\mu\lambda\nu\sigma}-g_{\mu\nu}R_{\lambda\sigma}\right)\frac{\nabla^\lambda\nabla^\sigma F_R}{F_R},\label{sigma-mn}\eea comprises the contribution coming from the fourth-order curvature terms, $\nabla^2\equiv g^{\mu\nu}\nabla_\mu\nabla_\nu$ and $T^{(m)}_{\mu\nu}$ is the stress-energy tensor of the matter degrees of freedom. While deriving the EOM \eqref{sigma-mn} we have taken into account the relationships \eqref{rel}. The trace of \eqref{feq}: 

\bea 3\frac{\nabla^2F_R}{F_R}-\frac{4\beta}{\alpha}G_{\lambda\sigma}\frac{\nabla^\lambda\nabla^\sigma F_R}{F_R}+R+\frac{2\beta}{\alpha}{\cal G}-\frac{2F}{F_R}=\frac{\kappa^2}{F_R}\left[T^{(m)}+4\epsilon\mu^4\right],\label{trace}\eea where $T^{(m)}=g^{\mu\nu}T^{(m)}_{\mu\nu}$ is the trace of the stress-energy tensor of matter, amounts to an additional dynamical equation on the variable $F_R$.

%================================================================

\subsection{Small curvature limit ($R\approx{\cal G}\approx 0$)}

%================================================================

Let us to expand the function $F(R,{\cal G})$ at small curvature up to fourth-order curvature terms \cite{comelli}:

\bea F(R,{\cal G})=F_0+F^0_R R+\frac{1}{2}F^0_{RR}R^2+F^0_{\cal G}{\cal G},\label{sm-curv-expand}\eea where $F_0=F(R,{\cal G})=F(0,0)$, $F^0_R=\der F/\der R|_{(0,0)}$, etc. The following effective (low curvature) action is retrieved:

\bea S_\text{eff}=\frac{M^2_\text{Pl}}{2}\int d^4x\sqrt{-g}\left(R-2\Lambda_\text{eff}+\frac{1}{6m^2_0}R^2+\frac{\beta}{\alpha}{\cal G}\right),\label{eff-action}\eea where

\bea M^2_\text{Pl}=\frac{F^0_R}{\kappa^2},\;\Lambda_\text{eff}=-\frac{F_0}{2F^0_R}-\frac{\epsilon\mu^4}{M^2_\text{Pl}},\;m^2_0=\frac{F^0_R}{3F^0_{RR}}.\label{eff-quantities}\eea The Gauss-Bonnet term in \eqref{eff-action} amounts to a total divergence so that it does not modify the EOM and may be safely omitted.

The exchange of the extra scalar degree of freedom with mass $m_0$ between two test particles with masses $m_1$ and $m_2$, modifies the Newtonian gravitational potential through an additional Yukawa interaction:

\bea V(r)=-G_N\frac{m_1m_2}{r}\left[1+\alpha\exp\left(-\frac{m_0c}{\hbar}r\right)\right],\nonumber\eea where $c$ is the speed of light, $\hbar$ is the reduced Planck constant and

\bea m_0=\frac{\hbar}{c\lambda}=\frac{1.967}{\lambda}\times 10^{-10}\,\text{GeV},\label{m0-bound}\eea with the length scale $\lambda$ in $\mu$m. According to \cite{newton-law} the gravitational-strength Yukawa interactions are limited to ranges $\lambda<38.6\,\mu$m with $95\%$ confidence, so that $m_0>5\times 10^{-30}\,M_\text{Pl}$. Hence the following bound is to be satisfied:

\bea m^2_0=\frac{F^0_R}{3F^0_{RR}}>2.5\times 10^{-59}\;M^2_\text{Pl}.\label{newton-law-bond}\eea

It should be stressed that the effective action \eqref{eff-action}, which coincides with the one for the Starobinsky model \cite{staro-plb-1980, staro-plb-1982, vilenkin-1985, staro-jetp-2007, kehagias-2014} with a non-vanishing cosmological constant, is correct only for small curvatures down to scales of the order: 

\bea \frac{R^2}{6m^2_0}\sim R\ll\frac{1}{\alpha}\;\Rightarrow\;R^2\ll\frac{6m^2_0}{\alpha},\;\;{\cal G}\sim\frac{\alpha}{\beta}R\ll\frac{1}{\beta}.\label{sm-c-lim}\eea For much smaller curvature the effective action just coincides with the Einstein-Hilbert action:

\bea S^\text{EH}_\text{eff}=\frac{M^2_\text{Pl}}{2}\int d^4x\sqrt{-g}\left(R-2\Lambda_\text{eff}\right),\label{eh-action}\eea since, as $R\rightarrow 0$, the related curvature quantities $R^2$ and ${\cal G}$ vanish faster than $R$.

%%%%%%%%%%%%%%%%%%%%%%%%%%%%%%%%%%%%%%%%%%%%%%%

\section{Ghost freedom}\label{sect-ghost-free}

%%%%%%%%%%%%%%%%%%%%%%%%%%%%%%%%%%%%%%%%%%%%%%%

In order to investigate the propagating degrees of freedom, we shall study the linearization of the action \eqref{frg-action} around maximally symmetric spaces of constant curvature $R_0$ \cite{prd-2010}. In this case ${\cal G}_0=R_0^2/6$. We shall expand the action up to terms quadratic in the curvature, so that terms like $(R-R_0)^3$, $(R-R_0)\left({\cal G}-{\cal G}_0\right)$ and higher, will be omitted. We have that:

\bea F(R,{\cal G})=\tilde{F}_0+\tilde{F}_R^0\left(R-R_0\right)+\frac{1}{2}\tilde{F}_{RR}^0\left(R-R_0\right)^2+\tilde{F}_{\cal G}^0\left({\cal G}-{\cal G}_0\right)+{\cal O}(3),\nonumber\eea where $\tilde{F}_0\equiv F(R_0,{\cal G}_0)$, $\tilde{F}_R^0\equiv F_R(R_0,{\cal G}_0)$, etc. If reorganize the above equation we can write it in more compact form (in the given approximation):

\bea F(R,{\cal G})=\xi_0+\zeta_0R+\upsilon_0R^2+\omega_0{\cal G},\label{f-expand}\eea where we have introduced the following identifications:

\bea &&\xi_0\equiv\tilde{F}_0-R_0\tilde{F}_R^0+\frac{1}{2}R_0^2\tilde{F}_{RR}^0-\frac{1}{6}R_0^2\tilde{F}_{\cal G}^0,\nonumber\\
&&\zeta_0\equiv\tilde{F}_R^0-R_0\tilde{F}^0_{RR},\;\upsilon_0\equiv\frac{\tilde{F}_{RR}^0}{2},\;\omega_0\equiv\tilde{F}_{\cal G}^0.\nonumber\eea If substitute the above expansion back into the action \eqref{frg-action} we get:

\bea S=\frac{M^2_\text{Pl}}{2}\int d^4x\sqrt{-g}\left(R-2\Lambda+\frac{1}{6m^2}R^2\right),\label{action-expand-r0}\eea where

\bea M^2_\text{Pl}=\frac{\zeta_0}{\kappa^2},\;\Lambda=-\frac{\xi_0+2\epsilon\kappa^2\mu^4}{2\zeta_0},\;m^2=\frac{\zeta_0}{6\upsilon_0}=\frac{\tilde{F}_R^0-R_0\tilde{F}_{RR}^0}{3\tilde{F}_{RR}^0},\label{l-m}\eea and the term under the integral $\propto{\cal G}$ has been omitted since it amounts to a total derivative. It is a well-known fact that the linearization \eqref{action-expand-r0} is associated with three propagating degrees of freedom \cite{stelle, hindawi}: the two polarizations of the (massless) graviton and a massive scalar mode with mass squared $m^2$. In order to avoid a tachyon instability it is then required that:

\bea m^2\geq 0\;\Rightarrow\;\tilde{F}^0_R>R_0\tilde{F}^0_{RR},\;\tilde{F}^0_{RR}>0.\label{tachy-bond}\eea

As it was for the expansion around flat space, the present linearization is correct for small departures from de Sitter space with constant curvature $R_0$, down to the scale $R\sim m^2$. For much smaller curvature scales $R\ll m^2$, the action \eqref{action-expand-r0} reduces to the Einstein-Hilbert action.

%==============================================================================

\subsection{Ghosts due to anisotropy of space}\label{subsect-ghost-anisotropy}

%==============================================================================

Even if the $F(R,{\cal G})$ modified theory of gravity is free of ghosts when linearized around maximally symmetric spaces, when other less symmetric backgrounds such as anisotropic spaces, are considered, it is not for granted that the theory will be free of ghost in this latter case. In Ref. \cite{defelice-ptp-2010} the study of linear perturbation theory for general $F(R,{\cal G})$ was carried out over an empty anisotropic background of the Kasner-type:

\bea ds^2=-dt^2+a^2(t)dx^2+b^2(t)(dy^2+dz^2),\label{kasner}\eea where $a(t)$ and $b(t)$ are the scale factors, in order to show that, within general $F(R,{\cal G})$ theories, an anisotropic background has ghost degrees of freedom, which are absent on Friedmann-Robertson-Walker (FRW) backgrounds. Their study revealed that on this background the number of independent propagating degrees of freedom is four. It reduces to three on FRW backgrounds, since one mode becomes highly massive and decouples from the physical spectrum. The ghost mode is inevitable unless the following condition is fulfilled \cite{defelice-ptp-2010}:

\bea \left|\frac{\der(\chi,\xi)}{\der(R,{\cal G})}\right|=F_{RR}F_{\cal GG}-\left(F_{R{\cal G}}\right)^2=0,\label{no-ghost}\eea where $\chi$, $\xi$ are auxiliary fields introduced in the study of \cite{defelice-ptp-2010}. If the above condition is fulfilled, then perturbations of $\chi$ and $\xi$ are not independent. In general backgrounds this is true if \cite{defelice-ptp-2010} ${\cal L}=\chi(\phi)R+\xi(\phi){\cal G}-V(\phi)$. But this is not the only possibility left to avoid the ghosts due to anisotropy.

Actually, for theories of the kind we consider in this paper: $F(R,{\cal G})=F(R+c{\cal G})$, where $c=\beta/\alpha$ is a free constant, the condition \eqref{no-ghost} is fulfilled since, for this class of theories: $F_{\cal G}=cF_R$ and $F_{\cal GG}=cF_{R{\cal G}}=cF_{{\cal G}R}=c^2F_{RR}$, as seen from \eqref{rel}. Hence, for the latter more general class of fourth-order theories, ghosts due to anisotropy of space are absent.

%========================================================================================================

\subsection{Scalar perturbations and a modification of the dispersion relation}\label{subsect-kgroup-vel}

%========================================================================================================

By the same reason as above, i. e., that the relationships \eqref{rel} take place, neither a strong instability nor superluminal propagation occurs due to a modification of the dispersion relation found in \cite{defelice-jcap-2009}. In this reference the authors perform a general study of cosmological perturbations in vacuum for general $F(R,{\cal G})$ theories. They found a modification of the dispersion relation for scalar perturbations, in comparison with previous similar studies \cite{cartier, noh}, that leads to unwanted  -- either unstable or tachyonic -- behavior. Actually, in \cite{defelice-jcap-2009} the following non-standard wave equation was obtained for the gauge invariant (Fourier) field $\Phi$:

\bea \frac{1}{a^3Q}\der_t\left(a^3Q\dot\Phi\right)+B_1\frac{k^2}{a^2}\Phi+B_2\frac{k^4}{a^4}\Phi=0,\label{wave-eq}\eea where $Q=Q(t)$, $B_1=B_1(t)$ and $B_2=B_2(t)$ are time-dependent parameters and $k$ is the wavenumber of the perturbation. In this study the degrees of freedom $\Phi$ and $\dot\Phi$ are enough to describe the behavior of the metric perturbations. The above wave equation equation contains a term proportional to $k^4$, which does not vanish in generic $F(R,{\cal G})$ theories.\footnote{This term corresponds to fourth order spatial derivative in real space and is not a spurious result due to a bad choice of gauge since $\Phi$ is gauge invariant.} This term is responsible for non-standard behavior of the scalar metric perturbations. For instance, if $B_2$ were negative, then the Friedmann-Robertson-Walker (FRW) space were unstable on small scales (short wavelength limit) \cite{defelice-jcap-2009}. If $B_2$ were positive instead, up to the leading term the group velocity $v_g(k)\approx 2\sqrt{B_2}k/a$. It exceeds the speed of light for modes above the critical wavenumber $k_c=a/2\sqrt{B_2}$. Hence, the propagation of short wavelength modes eventually (inevitably) becomes superluminal. This is true, except for the above mentioned special cases where (equation (6.18) of Ref. \cite{defelice-jcap-2009}):

\bea F_{RR}F_{\cal GG}-F_{R{\cal G}}F_{{\cal G}R}=F_{RR}F_{\cal GG}-\left(F_{R{\cal G}}\right)^2=0.\label{except}\eea The small wavelength modes inevitably either suffer from strong instability or undergo superluminal propagation. 

In the kind of theories we are investigating here \eqref{frg-gen}, thanks to the relationship \eqref{rel}, the condition for absence of instability/tachyonic behavior \eqref{except}, is identically fulfilled. This means that the above discussed kind of instability is not present in the models of our interest.

%============================================

\subsection{Absence of other instabilities}

%============================================

Among the most dangerous instabilities, when higher-curvature corrections of gravity are considered, is the so called Dolgov-Kawasaki (matter) instability \cite{dolgov-kawasaki, nojiri-odintsov, faraoni-2006, sotiriou-faraoni}. This instability, which is specially important in the $f(R)$ theories since the curvature scalar $R$ is a dynamical degree of freedom \cite{faraoni-2006, sotiriou-faraoni}, is of special importance in the present setup as well. The stability criterion in this case requires that

\bea \frac{d\kappa^2_\text{eff}}{dR}=-\frac{\kappa^2F_{RR}}{F_R^2}<0.\label{dk-bound}\eea Hence, the Dolgov-Kawasaki instability is avoided only for non-negative $F_{RR}\geq 0$. If the $R$-derivative of $\kappa^2_\text{eff}$ in \eqref{dk-bound} were positive, the effective gravitational coupling increased with the curvature, so that, at larger curvature gravity becomes stronger which then implies that $R$ itself generates a larger curvature through the trace equation \eqref{trace}. In other words, a positive feedback mechanism acts to destabilize the theory \cite{prd-2010, sotiriou-faraoni}. 

In addition to the above stability criteria, a constraint coming from requiring positivity of the effective gravitational coupling: 

\bea \kappa^2_\text{eff}=\frac{\kappa^2}{F_R}>0,\label{k2-eff-pos}\eea is also to be satisfied. 

As shown in \cite{defelice-jcap-2009}, for models of the class $F(R,{\cal G})=F(R+c{\cal G})$, like the ones we are interested in here, the wave equation \eqref{wave-eq} for the scalar perturbations is given by:

\bea \frac{1}{a^3Q}\der_t\left(a^3Q\dot\Phi\right)+c^2_s\frac{k^2}{a^2}\Phi=0,\label{wave-eq-frg}\eea where the squared sound speed is defined in the following way:

\bea c^2_s=1+\frac{8\beta\dot H/\alpha}{1+4\beta H^2/\alpha}.\nonumber\eea We should require non-negative squared sound speed $c^2_s\geq 0$ since, otherwise, a Laplacian or gradient instability develops. Meanwhile, for any $F(R,{\cal G})$ model, for the squared speed of propagation of the tensor modes one gets \cite{defelice-jcap-2009}:

\bea c^2_T=\frac{F_R+4\beta\ddot F_R/\alpha}{F_R+4\beta H\dot F_R/\alpha}.\nonumber\eea The absence of ghosts requires that $F_R+4\beta H\dot F_R/\alpha>0$, while, in order to avoid the Laplacian instability: $c^2_T\geq 0$.

The absence of ghosts and instabilities such as: ghosts due to anisotropy of space or to linear perturbations around spherically symmetric static background \cite{spheric-perts}, tachyonic, Dolgov-Kawasaki and Laplacian instabilities, is required if the given $F(R,{\cal G})$ theories are phenomenologically viable options for the description of our universe. In particular, the choice in \eqref{frg-gen} makes these theories very attractive possibilities for viable fourth-order theories of gravity since all of the mentioned instabilities may be avoided in a given subspace of the parameter's space.

%%%%%%%%%%%%%%%%%%%%%%%%%%%%%%%%%%%%%%%%%%%%%%%%%%%%%%%%%%%%%%%%%%%%%%%%%%%%%%%%%%

\section{Power-law $F(R+c{\cal G})$ models of modified gravity}\label{sect-models}

%%%%%%%%%%%%%%%%%%%%%%%%%%%%%%%%%%%%%%%%%%%%%%%%%%%%%%%%%%%%%%%%%%%%%%%%%%%%%%%%%%

Here we study a three-parametric class of models of the kind $F(R+c{\cal G})$ modified gravity and check them to stability and phenomenological viability. For the present choice of the $F(R,{\cal G})$ modification of gravity, several sources of instability such as ghosts due to anisotropy of space and non-standard behavior of the scalar metric perturbations -- potentially leading to superluminal propagation of short wavelength modes -- are eliminated. However, avoidance of other kinds of instability such as the Dolgov-Kawasaki and Laplacian instabilities, as well as the requirement of positivity of the effective gravitational coupling, lead to additional constraints on the parameter space. 

In the present case we choose the following power-law function $F(R,{\cal G})$:

\bea F(R,{\cal G})=-\lambda^2\left(1-\alpha R-\beta{\cal G}\right)^\nu,\label{pwl-frg}\eea where $\lambda$, $\alpha$ and $\beta$ are free constants with mass dimensions $M$, $M^{-2}$ and $M^{-4}$, respectively, while $\nu$ is a dimensionless constant. Notice that, although there are four free parameters in \eqref{pwl-frg}, the parameter $\lambda^2$ may be combined with $\kappa^2$ in \eqref{frg-action}, so that the resulting $F(R,{\cal G})$ is actually a three-parametric function. We have that:

\bea &&F_R=\alpha\lambda^2\nu\left(1-\alpha R-\beta{\cal G}\right)^{\nu-1},\;F_{\cal G}=\beta\lambda^2\nu\left(1-\alpha R-\beta{\cal G}\right)^{\nu-1},\nonumber\\
&&F_{R{\cal G}}=F_{{\cal G}R}=-\alpha\beta\lambda^2\nu(\nu-1)\left(1-\alpha R-\beta{\cal G}\right)^{\nu-2},\nonumber\\
&&F_{RR}=-\alpha^2\lambda^2\nu(\nu-1)\left(1-\alpha R-\beta{\cal G}\right)^{\nu-2},\;F_{{\cal G}{\cal G}}=-\beta^2\lambda^2\nu(\nu-1)\left(1-\alpha R-\beta{\cal G}\right)^{\nu-2},\label{f-der}\eea so that the relationships \eqref{rel} are satisfied. In what follows we shall assume that the following constraint on the curvature quantities is satisfied:

\bea 1-\alpha R-\beta{\cal G}\geq 0.\label{const-frg}\eea Under the above assumption, for the three-parametric class of function \eqref{pwl-frg}, absence of the Dolgov-Kawasaki instability and positivity of the effective gravitational coupling -- requirements \eqref{dk-bound} and \eqref{k2-eff-pos}, respectively -- amount to the following conditions:

\bea &&\frac{d\kappa^2_\text{eff}}{dR}=\frac{\kappa^2(\nu-1)\left(1-\alpha R-\beta{\cal G}\right)^{-\nu}}{\lambda^2\nu}<0,\nonumber\\
&&\kappa^2_\text{eff}=\frac{\kappa^2\left(1-\alpha R-\beta{\cal G}\right)^{1-\nu}}{\alpha\lambda^2\nu}>0.\label{cond-1}\eea Hence, phenomenologically viable theories of this type require that $\alpha>0$ and $0<\nu<1$.

The so called Born-Infeld inspired $F(R,{\cal G})$ models of the kind \cite{comelli, jcap-2010}:

\bea F(R,{\cal G})=-\lambda^2\sqrt{1-\alpha R-\beta{\cal G}},\label{bi-frg}\eea fall into the above phenomenologically viable class of models of modified gravity, when we set $\nu=1/2$. In what follows we shall focus in the investigation, specifically, of this two-parametric class of models.

%%%%%%%%%%%%%%%%%%%%%%%%%%%%%%%%%%%%%%%%%%%%%%%%%%%%%%%%%%%%%%%%%%%%%%%%%%%%%%%%%%%%%%%%%%%

\section{Asymptotic dynamics of BI-inspired $F(R,{\cal G})$ cosmology}\label{sect-bigb-2d}

%%%%%%%%%%%%%%%%%%%%%%%%%%%%%%%%%%%%%%%%%%%%%%%%%%%%%%%%%%%%%%%%%%%%%%%%%%%%%%%%%%%%%%%%%%%

Here we shall investigate the cosmological dynamics of the BI-inspired $F(R,{\cal G})$ model \eqref{bi-frg} with action \eqref{frg-action}, in a FRW background space with flat spatial sections, whose line-element reads: 

\bea ds^2=-dt^2+a^2(t)\delta_{ik}dx^idx^j,\;\;i,j=1,2,3.\label{frw-le}\eea In this case we have that:

\bea R=6\dot H+12H^2,\;{\cal G}=24H^2\left(\dot H+H^2\right),\label{r-g-frw}\eea where $H=\dot a/a$ is the Hubble parameter. 

For the $F(\alpha R+\beta{\cal G})$ class of function the FRW equations of motion \eqref{feq} read:

\bea &&3H^2+3H\frac{\dot F_R}{F_R}\left(1+4\frac{\beta}{\alpha}H^2\right)-\frac{1}{2}\left(R+\frac{\beta}{\alpha}{\cal G}-\frac{F}{F_R}\right)=\frac{\kappa^2}{F_R}\left(\rho_m-\epsilon\mu^4\right),\label{fried}\\
&&\frac{\ddot F_R}{F_R}=-\frac{\kappa^2(\omega_m+1)\rho_m}{F_R\left(1+4\frac{\beta}{\alpha}H^2\right)}+H\frac{\dot F_R}{F_R}-\frac{2\left(1+4\frac{\beta}{\alpha}H\frac{\dot F_R}{F_R}\right)}{1+4\frac{\beta}{\alpha}H^2}\dot H,\label{raycha}\eea where $\rho_m$ and $p_m=\omega_m\rho_m$ are the energy density and pressure of the matter fluid, while $\omega_m$ is its equation of state (EOS) parameter, respectively. In the present case the trace equation \eqref{trace} is not an independent equation so that we do not write it. The above EOM-s can be written in the following alternative way:

\bea &&\frac{\dot H}{H^2}=-1+\frac{\Omega_m-\Omega_{\mu^4}-2}{1+4\frac{\beta}{\alpha}H^2}-\frac{\dot F_R}{HF_R}+\frac{1}{3\alpha H^2(1+4\frac{\beta}{\alpha}H^2)},\label{hdot}\\
&&\frac{\ddot F_R}{H^2F_R}=-\frac{3(\omega_m+1)\Omega_m}{1+4\frac{\beta}{\alpha}H^2}+\frac{\dot F_R}{HF_R}-\frac{2\left[1+4\frac{\beta}{\alpha}H^2\left(\frac{\dot F_R}{HF_R}\right)\right]}{1+4\frac{\beta}{\alpha}H^2}\frac{\dot H}{H^2},\label{fddot}\eea where we have introduced the dimensionless energy densities of matter and of $\mu^4$:

\bea \Omega_m\equiv\frac{\kappa^2\rho_m}{3F_RH^2},\;\Omega_{\mu^4}\equiv\frac{\epsilon\kappa^2\mu^4}{3F_RH^2}.\label{dim-density}\eea Notice that, for $\epsilon=+1$ the constant $\mu^4$ contributes a negative energy density. This is not problematic since the effective cosmological constant at low curvature is 

\bea \Lambda_\text{eff}=\frac{\lambda^2-2\epsilon\kappa^2\mu^4}{\alpha\lambda^2},\label{eff-cc}\eea as seen from \eqref{eff-quantities}. Hence, as long as $\lambda^2\geq 2\epsilon\kappa^2\mu^4$, $\Lambda_\text{eff}$ is a non-negative quantity even if $\epsilon=+1$.

For the choice \eqref{bi-frg}, the assumption \eqref{const-frg} is not an independent requirement but a constraint on the physical viability of the resulting cosmological model. Hence, for the specific model of interest here, this constraint amounts to a phenomenological bond which, in FRW space, can be written in the following way:

\bea \frac{\dot H}{H^2}\leq\frac{1}{6\alpha H^2\left(1+4\frac{\beta}{\alpha}H^2\right)}-\frac{2\left(1+2\frac{\beta}{\alpha}H^2\right)}{1+4\frac{\beta}{\alpha}H^2}.\label{phenom-viab}\eea

%========================================================================================================

\subsection{Simplified dynamical system: matter vacuum with vanishing vacuum energy}\label{subsect-0vac}

%========================================================================================================

In what follows, for simplicity, we shall investigate the particular case when the density of matter vanishes $\Omega_m=0$ (vacuum case) and $\mu^4=0$. This means that at small curvature:

\bea S=\frac{M^2_\text{Pl}}{2}\int d^4x\sqrt{-g}\left(R-\frac{2}{\alpha}+\frac{\alpha}{4}R^2\right),\label{small-c}\eea where $M^2_\text{Pl}=\alpha\lambda^2/2\kappa^2$. Although this is not the most general situation in which we may have even a vanishing effective cosmological constant, anyway the basic features of the model are preserved.\footnote{Recall that for non-vanishing $\mu^4$, the cosmological constant at small curvature: $\Lambda_\text{eff}=(\lambda^2-2\epsilon\kappa^2\mu^2)/\alpha\lambda^2$, can be made as small as one desires by properly arranging the parameters $\lambda^2$ and $\mu^2$ if $\epsilon=+1$. For instance, by letting $\lambda^2=2\kappa^2\mu^2+\delta\lambda$, where $\delta\lambda$ is a very small quantity.}  

%------------------------------------------------------------

In order to perform the asymptotic dynamics analysis of this model, let us introduce the following dimensionless (bounded) variables of some state space:\footnote{For a compact introduction to the dynamical systems analysis close to this presentation see \cite{quiros-ejp-2015}.}

\bea &&x=\frac{1}{1+4\frac{\beta}{\alpha}H^2}\;\Rightarrow\;4\frac{\beta}{\alpha}H^2=\frac{1-x}{x},\;0\leq x\leq 1,\nonumber\\
&&y_\pm\equiv\frac{\dot F_R}{HF_R\pm\dot F_R}\;\Rightarrow\;\left[\frac{\dot F_R}{HF_R}\right]_\pm=\frac{y_\pm}{1\mp y_\pm},\;-1\leq y_-\leq 0,\;0\leq y_+\leq 1,\label{vars}\eea where the whole phase space is covered by the bounded variables $x\in [0,1]$ and $y=y_-\cup y_+\in [-1,1]$. The cosmological equations \eqref{fried}, \eqref{raycha} for the case of interest can be traded, accordingly, by the following autonomous dynamical system on these variables:

\bea &&x'=-2x(1-x)\left[\frac{\dot H}{H^2}\right]_\pm,\nonumber\\
&&y'_\pm=(1\mp y_\pm)^2\left[\frac{\ddot F_R}{H^2F_R}\right]_\pm-y_\pm(1\mp y_\pm)\left[\frac{\dot H}{H^2}\right]_\pm-y_\pm^2,\label{ode}\eea where 

\bea &&\left[\frac{\dot H}{H^2}\right]_\pm=-1-2x-\frac{y_\pm}{1\mp y_\pm}+\frac{4\beta x^2}{3\alpha^2(1-x)},\nonumber\\
&&\left[\frac{\ddot F_R}{H^2F_R}\right]_\pm=\frac{y_\pm}{1\mp y_\pm}-\frac{2[x(1\mp y_\pm)+(1-x)y_\pm]}{1\mp y_\pm}\left[\frac{\dot H}{H^2}\right]_\pm,\label{hdot-vars}\eea and the prime denotes derivative with respect to the time variable $N=\ln a$. Notice that there are two different dynamical systems in \eqref{ode}; one for the choice of the '$+$' sign and another one for the choice '$-$'. However, these describe a unique phase space spanned by the variables $x$ and $y=y_-\cup y_+$. 

The model \eqref{bi-frg} is phenomenologically viable only if the function $F=F(R,{\cal G})$ is a real quantity, i. e., if the condition \eqref{phenom-viab} is fulfilled. For the present case, in terms of the variables $x$, $y$ the latter reads:

\bea \left[\frac{\dot H}{H^2}\right]_\pm\leq-1-x+\frac{2\beta x^2}{3\alpha^2(1-x)},\label{phenom-viab-1}\eea or $y_\pm\geq y_*^\pm$, where

\bea y_*^-=\frac{-x(1-x)+\frac{2\beta}{3\alpha^2}x^2}{(1-x)(1+x)-\frac{2\beta}{3\alpha^2}x^2},\;y_*^+=\frac{-x(1-x)+\frac{2\beta}{3\alpha^2}x^2}{(1-x)^2+\frac{2\beta}{3\alpha^2}x^2}.\label{viable}\eea Hence, the physically meaningful phase space corresponds to the following region of the plane: $\Psi_\text{2D}=\Psi^-_\text{2D}\cup\Psi^+_\text{2D}$,

\bea \Psi^-_\text{2D}=\left\{(x,y):\;0\leq x\leq 1,\;-1\leq y\leq 0,\;y\geq y_*^-\right\},\;\Psi^+_\text{2D}=\left\{(x,y):\;0\leq x\leq 1,\;0\leq y\leq 1,\;y\geq y_*^+\right\}.\label{2d-ps}\eea From the first equation in \eqref{hdot-vars} it also follows that the expansion is accelerated ($q=-1-\dot H/H^2<0$) if $y<y_\dag^-$ for $-1\leq y^-\leq 0$, or if $y<y_\dag^+$ for $0\leq y^+\leq 1$, where:

\bea y_\dag^-=\frac{-2x(1-x)+\frac{4\beta}{3\alpha^2}x^2}{(1-x)(1+2x)-\frac{4\beta}{3\alpha^2}x^2},\;y_\dag^+=\frac{-2x(1-x)+\frac{4\beta}{3\alpha^2}x^2}{(1-x)(1-2x)+\frac{4\beta}{3\alpha^2}x^2}.\label{accel}\eea 

\begin{figure*}[t!]
\includegraphics[width=5cm]{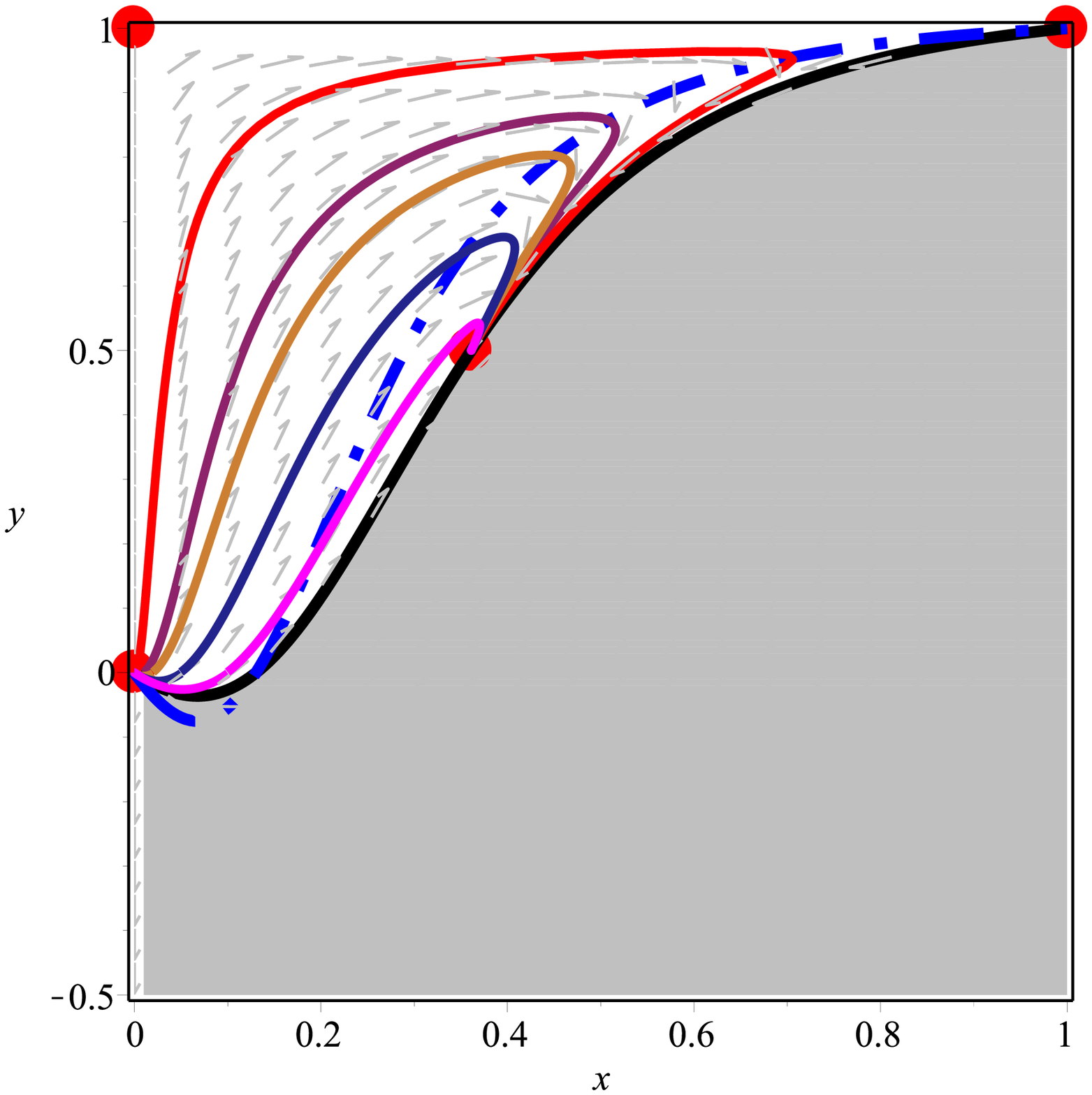}
\includegraphics[width=5cm]{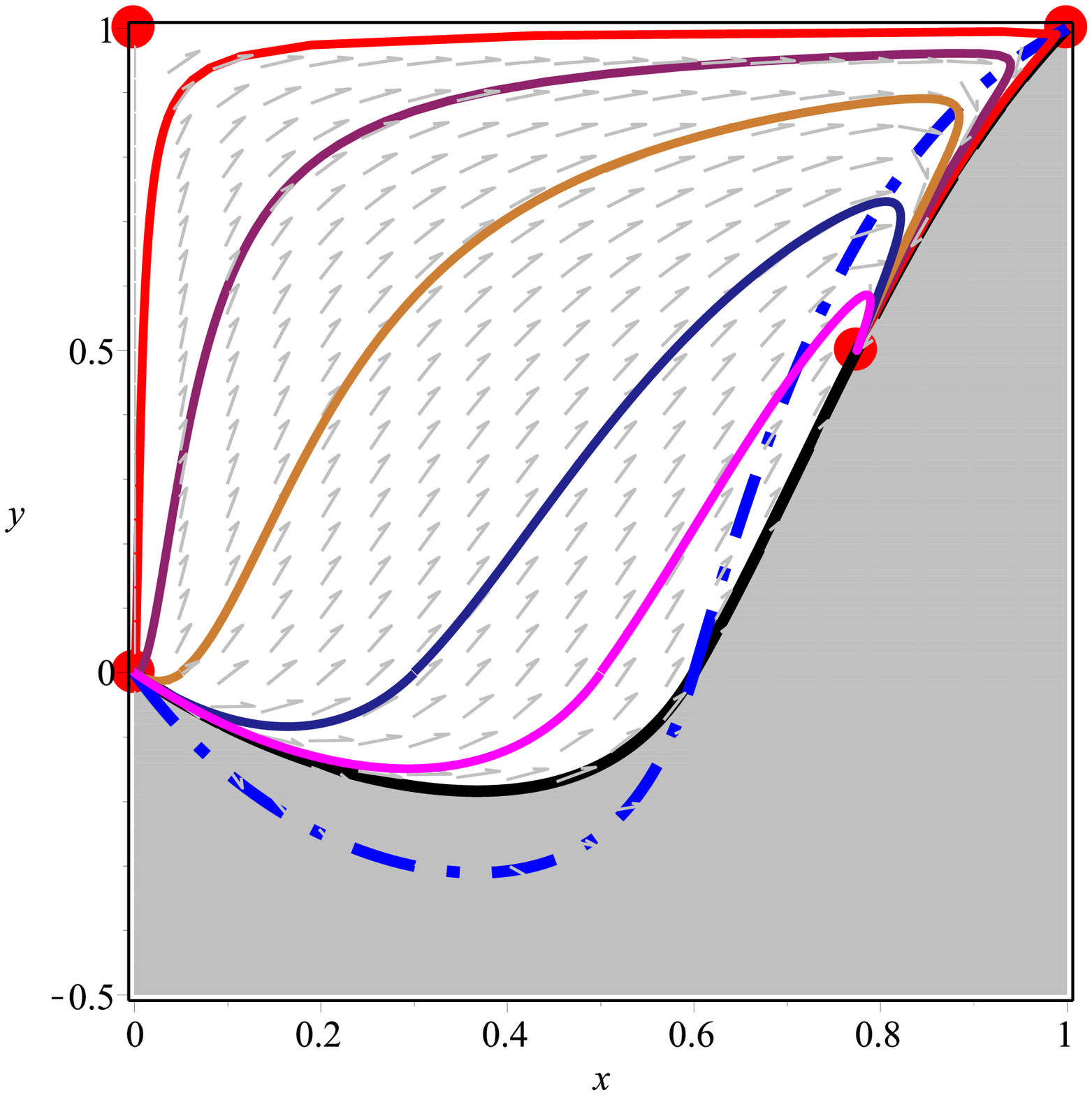}
\includegraphics[width=5cm]{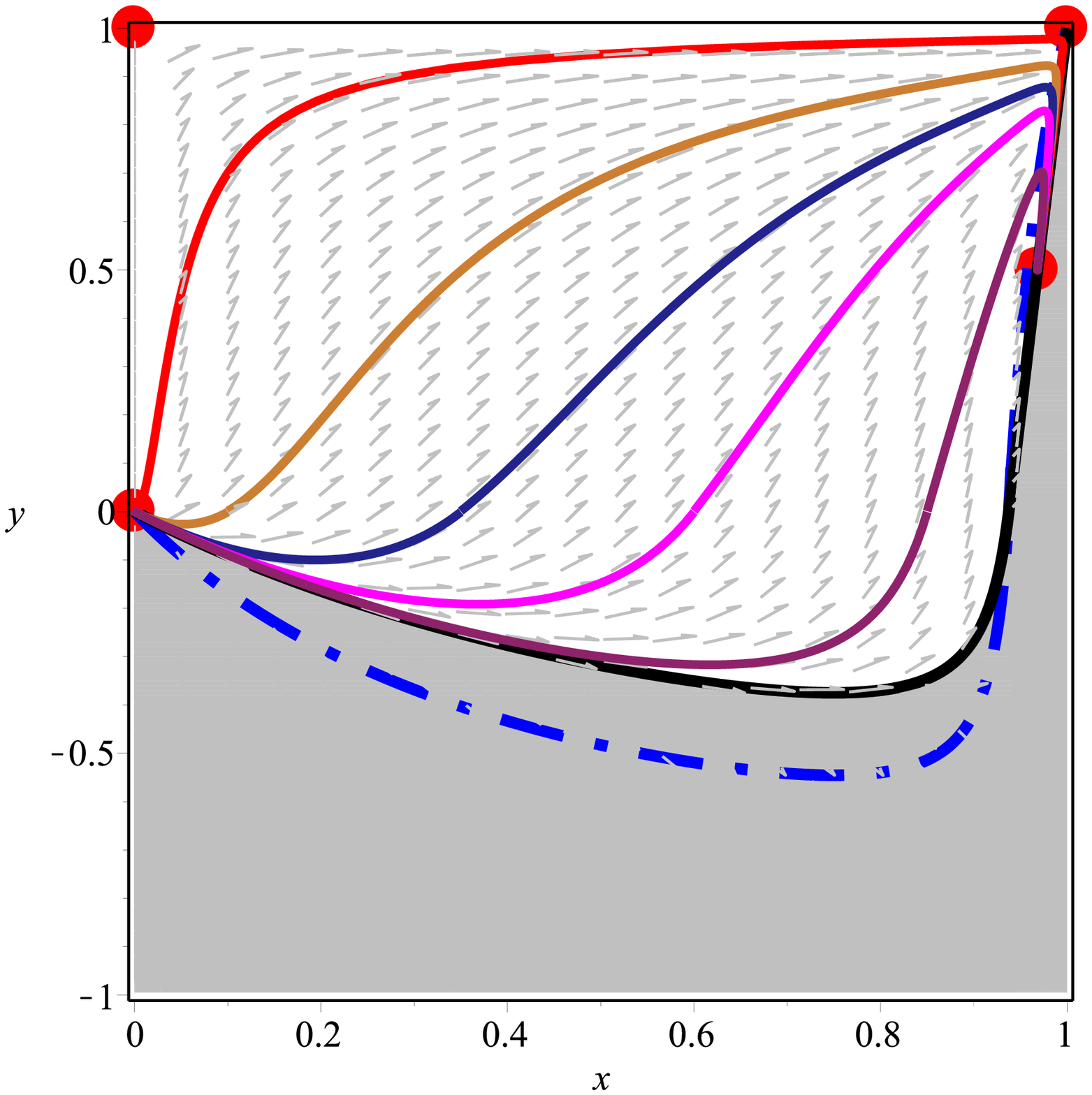}
\vspace{1.2cm}\caption{Phase portraits of the dynamical system \eqref{ode} for different (positive) values of the free parameters $\alpha$ and $\beta$. From left to the right $(\alpha,\beta)$: $(1,10)$, $(1,1)$ and $(1,0.1)$, so that the dimensionless ratio $\beta/\alpha^2$ equals $10$, $1$ and $10^{-1}$, respectively. The 'gray' region is unphysical since the condition \eqref{phenom-viab} is not satisfied. The critical points are represented by the small (red) solid circles. The thick dash-dot (blue) curve represents the condition $\dot H/H^2=-1$. Hence, the critical points that are located below this curve represent accelerated expansion.}\label{fig1}\end{figure*}

The phase portraits corresponding to the dynamical system \eqref{ode} are shown in FIG. \ref{fig1}, for different values of the free parameters $\alpha$ and $\beta$. The different orbits appearing in these phase portraits are generated by given sets of initial conditions $(x_i(0),y_i(0))$. Each orbit may be associated with a whole cosmic history, starting (possibly) in a past attractor (origin of the given evolutionary pattern) and ending up in a future attractor (destiny of the cosmic evolution). The 'gray' region is unphysical since the condition \eqref{phenom-viab} is not fulfilled. Hence, this region is excluded from the phase space $\Psi$. The thick dash-dot curves represent the condition $q=-1-\dot H/H^2=0$. In consequence, points in $\Psi$ that are located below these curves represent accelerated expansion. In what follows we consider only non-negative $\beta$-s. The parameter $\alpha$ is also positive due to the requirements of absence of the Dolgov-Kawasaki instability and of positivity of the gravitational coupling as we have discussed before. The case with negative Gauss-Bonnet coupling ($\beta<0$) has been studied in detail in \cite{jcap-2010}.

%-----------------------------------

%----------------------------------------

%---------------------------------------------------------------------

\subsection{Critical points of the dynamical system}\label{sub-cpoints}

Below we list those isolated critical points $P_i:(x_i,y_i)$ of the dynamical system \eqref{ode} in $\Psi_\text{2D}$, that are located within the phenomenologically viable region, together with their main properties.\footnote{In what follows, without loss of generality, we call a given point of the phase space as a ``critical point'' only if it is located within the phenomenologically viable region of the phase space, i. e., if it is in $\Psi_\text{2D}$, no matter whether it is, mathematically speaking, a critical point of the dynamical system.} These points are marked by small (red) solid circles in FIG. \ref{fig1}.

\begin{enumerate}

%---------------------------------------------------------------------------

\item{\it Origin}. The point $P_O:(0,0)$ is the global past attractor in the phase space $\Psi_\text{2D}$ since the eigenvalues of the linearization matrix for this point: $\lambda_1=2$ and $\lambda_2=4$, are both positive. At this point:

\bea x=0\;\Rightarrow\;H^2\gg\alpha/\beta,\;y=0\;\Rightarrow\;\frac{\dot F_R}{HF_R}=-\frac{\dot F}{HF}\rightarrow 0.\label{p4cond-1}\eea Besides, the function $F$ is undefined at this equilibrium point. Since at $P_O$, the deceleration parameter $q=-1-\dot H/H^2=0$, then:

\bea \frac{\dot H}{H^2}\rightarrow-1\;\;\Rightarrow\;\;H=t^{-1}\;\Rightarrow\;a\propto t.\label{p4cond-2}\eea This means that the evolution of the Universe starts in a big-bang singularity where $a(t)\rightarrow 0$ and $\dot H\approx-H^2\rightarrow-\infty$.

%------------------------------------------------------------------------------------

\item{\it Transient stages}.

\begin{itemize}

\item Point $P^0_\text{sdd}:(0,1)$ is a saddle critical point since the eigenvalues of the corresponding linearization matrix are of different sign. Hence, this point is associated with a transient state of the cosmic evolution. It is characterized by a very high curvature with $H^2\gg\alpha/\beta$ and $\dot F_R/HF_R$ undefined:

\bea y\rightarrow 1\;\Rightarrow\;\frac{\dot F_R}{HF_R}\rightarrow\infty\;\;(H>0),\label{p1cond-1}\eea which means, in turn, that $F_R\rightarrow 0$, i. e., that $F\rightarrow\infty$. This latter limit implies that, at least, 

\bea \frac{\dot H}{H^2}<-1\;\;\Rightarrow\;\;q>0\;(H>0),\label{p1cond-2}\eea i. e., this point represents a transient stage of decelerated expansion. As a matter of fact, since at $P^0_\text{sdd}$, $x=0$ and $y=1$, then (see first equation in \eqref{hdot-vars}):

\bea \frac{\dot H}{H^2}\rightarrow-\infty\;\;\Rightarrow\;\;q\rightarrow\infty.\label{p1cond-3}\eea This point should be associated with a curvature singularity.

\item The point $P^1_\text{sdd}:(1,1)$, where

\bea x=\frac{1}{1+4\frac{\beta}{\alpha}H^2}=1,\;y=1\;\Rightarrow\;\frac{\dot F_R}{HF_R}\rightarrow\pm\infty,\label{p5cond-1}\eea represents a transient cosmic stage with low curvature $H^2\ll\alpha/\beta$ (the numerical investigation reveals that this is a saddle equilibrium point as well). It is seen from the first equation in \eqref{hdot-vars} that at $P^1_\text{sdd}$, due to the competition between the negative (first) and the positive (last) terms, the quantity $\dot H/H^2$ is undefined. As a matter of fact this equilibrium state represents a turning point in what regards to the peace of the cosmic expansion: It is seen from FIG. \ref{fig1} that, as the given orbit evolves in the vicinity of $P^1_\text{sdd}$ in the route from the past into the future attractors, the cosmic history turns from decelerating into accelerating expansion. Recall that the curve \eqref{accel} corresponding to $q=0$, joints the points $P_O$ and $P^1_\text{sdd}$.

\end{itemize}

%-------------------------------------------------------------------------------------

\item{\it Destiny}. The point $$P_\text{dS}:\left(\frac{1}{\sqrt{1+2\beta/3\alpha^2}},\frac{1}{2}\right),$$ is a future attractor since the eigenvalues of the corresponding linearization matrix: $\lambda_1=-2$ and $\lambda_2=-4$, are both negative. It is the global future attractor in $\Psi_\text{2D}$. This critical point represents a de Sitter solution since $q=-1$ $\Rightarrow$ $\dot H=0$ $\Rightarrow$ $H=H_0$. It is a de Sitter attractor with constant Hubble rate squared:

\bea H^2_0=\frac{\alpha}{4\beta}\left(\sqrt{1+\frac{2\beta}{3\alpha^2}}-1\right).\label{p6cond-1}\eea

%--------------------------------------------------------------------------------------

\end{enumerate}

%%%%%%%%%%%%%%%%%%%%%%%%%%%%%%%%%%%%%%%%%%%%%%%%%%%%%%%%%%%%%%%%%%%%%%%%%%%%%%%%%%%%%%%%%%%%%%%%%

\section{Three dimensional phase space dynamics and the de Sitter solutions}\label{sect-bigb-3d}

%%%%%%%%%%%%%%%%%%%%%%%%%%%%%%%%%%%%%%%%%%%%%%%%%%%%%%%%%%%%%%%%%%%%%%%%%%%%%%%%%%%%%%%%%%%%%%%%%

In the above section we have investigated in detail the asymptotic dynamics of the BI inspired $F(R,{\cal G})$ theory \eqref{bi-frg} in the simplified case when $\mu^4=0$, i. e., vanishing vacuum energy density (we are investigating the vacuum case exclusively, i. e., $\Omega_m=0$). In this simplified case the asymptotic dynamics is described in a 2-dimensional (2D) phase space which means, in turn, that the mathematical handling is simpler. However, the assumption that $\mu^4=0$, means that at small curvature, where the effective theory is given by the action:

\bea S=\frac{M^2_\text{Pl}}{2}\int d^4x\sqrt{-g}\left(R-\frac{2}{\alpha}+\frac{\alpha}{4}R^2\right),\nonumber\eea we are not able to set the cosmological constant $1/\alpha$ to any negligible small value -- as required by the observations -- without forcing a unnaturally large coupling to the higher-curvature contribution. This is why, in the present section, we shall consider a non-vanishing vacuum energy $\mu^4\neq 0$. This amounts to increasing the dimension of the phase space from 2D to 3D. The corresponding mathematical handling is by far more complex. Our strategy to simplify the mathematics y to focus, exclusively, in the de Sitter solutions which are the ones that can be associated with the late-time accelerated expansion of the universe.

%==============================================================

\subsection{Dynamical system and the de Sitter critical points}

In the present case where $\mu^4\neq 0$, in addition to the phase space variables $x$, $y$ in \eqref{vars}, it is convenient to introduce the new bounded variable:

\bea u=\frac{\Omega_{\mu^4}}{\Omega_{\mu^4}+\epsilon}\;\Rightarrow\;\Omega_{\mu^4}=\frac{\epsilon u}{1-u},\label{u-var}\eea where $0\leq u\leq 1$ and $\epsilon=\pm 1$. The resulting 3D dynamical system reads:

\bea &&x'=-2x(1-x)\left[\frac{\dot H}{H^2}\right]_\pm,\nonumber\\
&&y'_\pm=(1\mp y_\pm)^2\left[\frac{\ddot F_R}{H^2F_R}\right]_\pm-y_\pm^2-y_\pm(1\mp y_\pm)\left[\frac{\dot H}{H^2}\right]_\pm,\nonumber\\
&&u'=-u(1-u)\left\{\frac{y_\pm}{1\mp y_\pm}+2\left[\frac{\dot H}{H^2}\right]_\pm\right\},\label{mu4-ode}\eea where, as before, the prime denotes derivative with respect to the time variable $N=\ln a$. Besides:

\bea &&\left[\frac{\dot H}{H^2}\right]_\pm=-1-\left[\frac{2+(\epsilon-2)u}{1-u}\right]x-\frac{y_\pm}{1\mp y_\pm}+\frac{4\beta x^2}{3\alpha^2(1-x)},\nonumber\\
&&\left[\frac{\ddot F_R}{H^2F_R}\right]_\pm=\frac{y_\pm}{1\mp y_\pm}-\frac{2[x(1\mp y_\pm)+(1-x)y_\pm]}{1\mp y_\pm}\left[\frac{\dot H}{H^2}\right]_\pm.\label{mu4-hdot}\eea 

In this section we shall focus in the de Sitter solutions exclusively. In order to check whether the corresponding critical points are within the phenomenologically viable region, it is required that $F^2\geq 0$, where $F$ is defined in \eqref{bi-frg}. Hence,

\bea \frac{\dot H}{H^2}\left(1+4\frac{\beta}{\alpha}H^2\right)+2+4\frac{\beta}{\alpha}H^2\leq\frac{1}{6\alpha H^2},\nonumber\eea but, since at the de Sitter points $\dot H=0$, then, for these critical points to be in the physically meaningful region it is required that:

\bea \frac{1}{\sqrt{1+2\beta/3\alpha^2}}\leq x\leq 1.\label{ineq}\eea 

At the de Sitter point, for $y\geq 0$, the equations \eqref{mu4-ode} and \eqref{mu4-hdot} become:

\bea &&x'=0,\;y'=y(1-2y),\;u'=\frac{u(1-u)y}{y-1},\label{mu4-ode0y+}\eea while for negative $y<0$: 

\bea &&x'=0,\;y'=y,\;u'=\frac{u(u-1)y}{y+1}.\label{mu4-ode0y-}\eea

%-----------------------------------------------------

\subsubsection{de Sitter critical manifold}

For $y=0$ one obtains a critical manifold:

\bea {\cal P}_\text{dS}=\left\{(x,0,u_*)\left|\;\frac{1}{\sqrt{1+2\beta/3\alpha^2}}\leq x\leq 1,\;u_*=u_*(x)\right\},\right.\label{cm-1}\eea where we have defined

\bea u_*(x)=\frac{(1-x)(1+2x)-\frac{4\beta}{3\alpha^2}x^2}{(1-x)[1+(2-\epsilon)x]-\frac{4\beta}{3\alpha^2}x^2}.\nonumber\eea Depending on location (smaller or larger $u$-values), points in ${\cal P}_\text{dS}$ can be either saddle critical points or local attractors instead. For points in ${\cal P}_\text{dS}$ we have that,

\bea 12\beta H^4+9\alpha H^2-\frac{F_R-\epsilon\alpha\kappa^2\mu^4}{F_R}=0.\label{usef-1}\eea On the other hand, since $y=0$ $\Rightarrow$ $\dot F_R=0$, one gets that $F_R=\tilde{F}^0_R=$const., where the value of the constant $\tilde{F}^0_R$ depends on the initial conditions.\footnote{This is a consequence of ${\cal P}_\text{dS}$ being a manifold instead of an isolated critical point.} From \eqref{usef-1} we obtain the following second-order algebraic equation:

\bea H^4+\frac{3\alpha}{4\beta}H^2-\frac{\tilde{F}^0_R-\epsilon\alpha\kappa^2\mu^4}{12\beta\tilde{F}^0_R}=0,\nonumber\eea whose real root is:

\bea H^2(F_0)=\frac{3\alpha}{8\beta}\left[\sqrt{1+\frac{16\beta}{27\alpha^2}\left(1-\frac{\epsilon\alpha\kappa^2\mu^4}{\tilde{F}^0_R}\right)}-1\right].\label{h2-mds}\eea

Given that for a set of points in ${\cal P}_\text{dS}$: the saddle points, the de Sitter solution is a transient stage, one may think that these can be associated with primordial inflation. However, as we shall see below, this is not the case since, if associate the late-time attractor $P_\text{dS}$ (see below) with the late-time inflationary stage, there is not possible to get the required amount of e-foldings of inflation. Hence, points in the above de Sitter manifold are to be associated with intermediate to late-time cosmological evolution.

%-----------------------------------------------------

\subsubsection{Isolated de Sitter attractor}

The other possibility is for $y=1/2$, where we are led with the isolated attractor:

\bea P^+_\text{dS}:\left(\frac{1}{\sqrt{1+2\beta/3\alpha^2}},\frac{1}{2},0\right).\label{mu4-attractor-ds}\eea At this point we have that \eqref{p6cond-1}:

\bea H^2_0=\frac{\alpha}{4\beta}\left(\sqrt{1+\frac{2\beta}{3\alpha^2}}-1\right).\nonumber\eea This isolated attractor is to be associated with the late-time stage of accelerated expansion of the universe. It can be checked that the ratio between $H^2(F_0)$ in \eqref{h2-mds} and $H^2_0$ above, at large $\alpha\gg 1$ amounts to:

\bea \frac{H^2(F_0)}{H^2_0}\approx\frac{4}{3}\left(1-\frac{\epsilon\alpha\kappa^2\mu^4}{\tilde{F}^0_R}\right).\nonumber\eea Hence, the saddle points in the manifold ${\cal P}_\text{dS}$ can not be associated with primordial inflation since there is not possible to get the required amount of inflation.

%-----------------------------------

\begin{figure*}[t!]
\includegraphics[width=5cm]{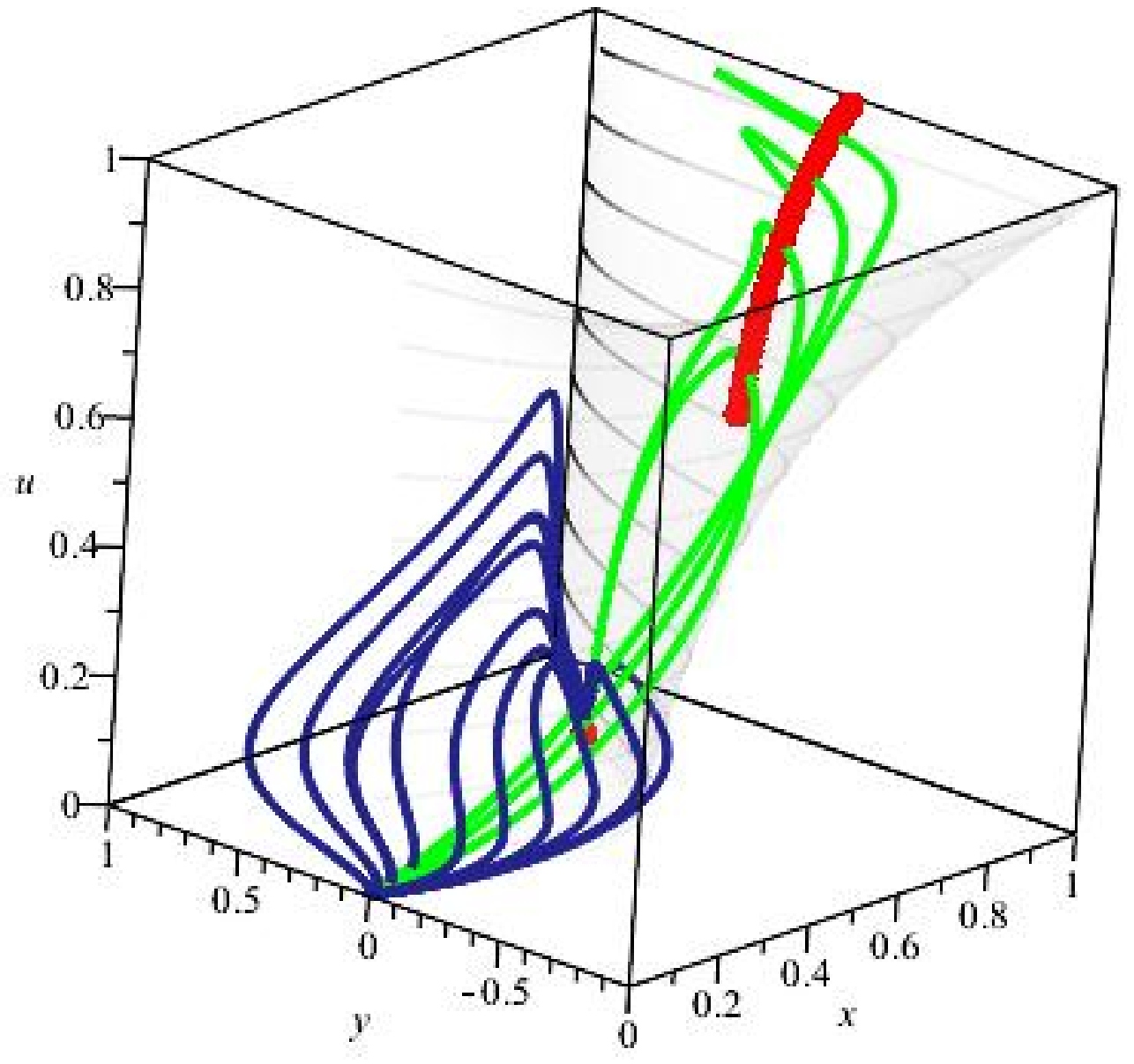}
\includegraphics[width=5cm]{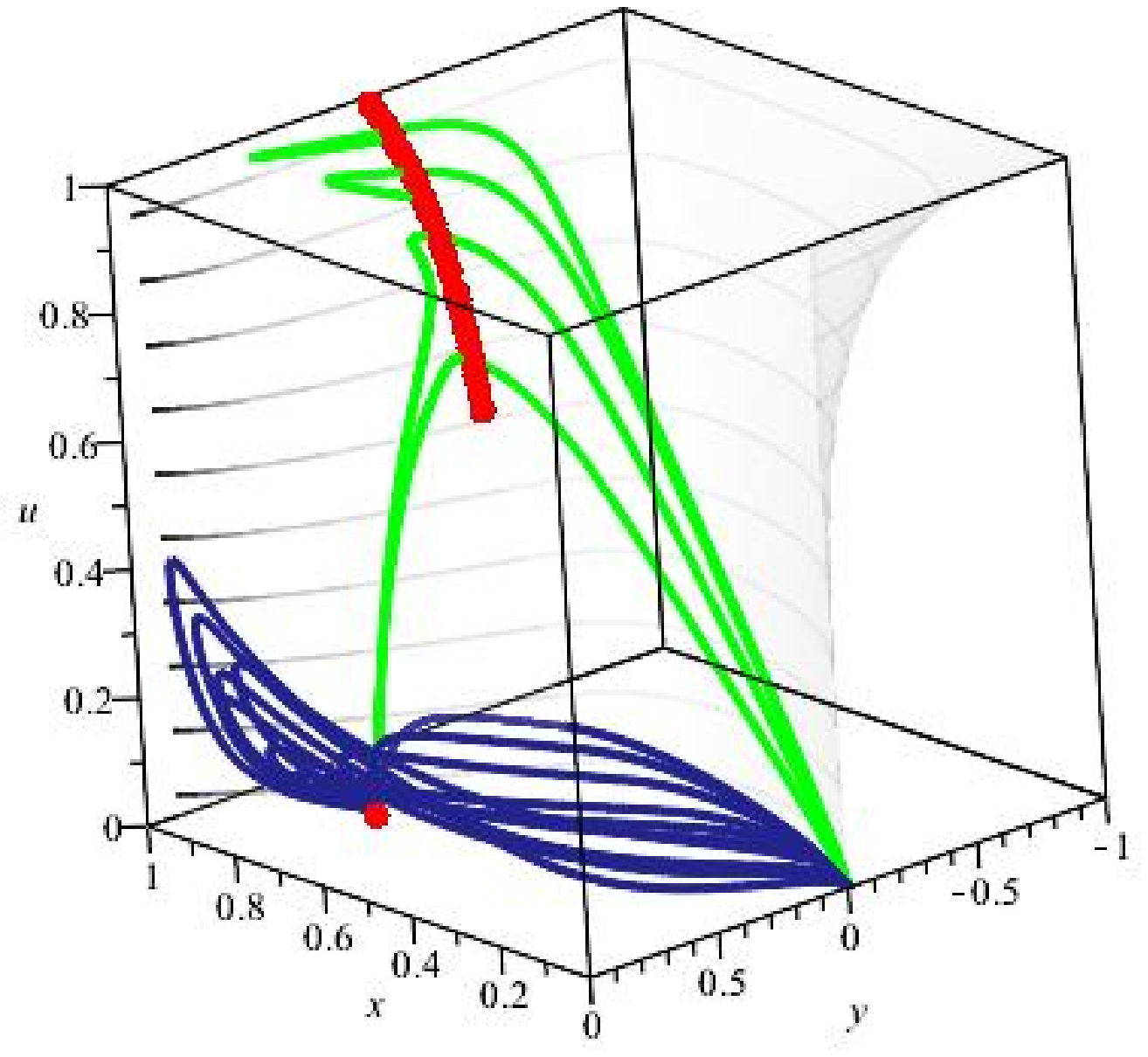}
\includegraphics[width=5cm]{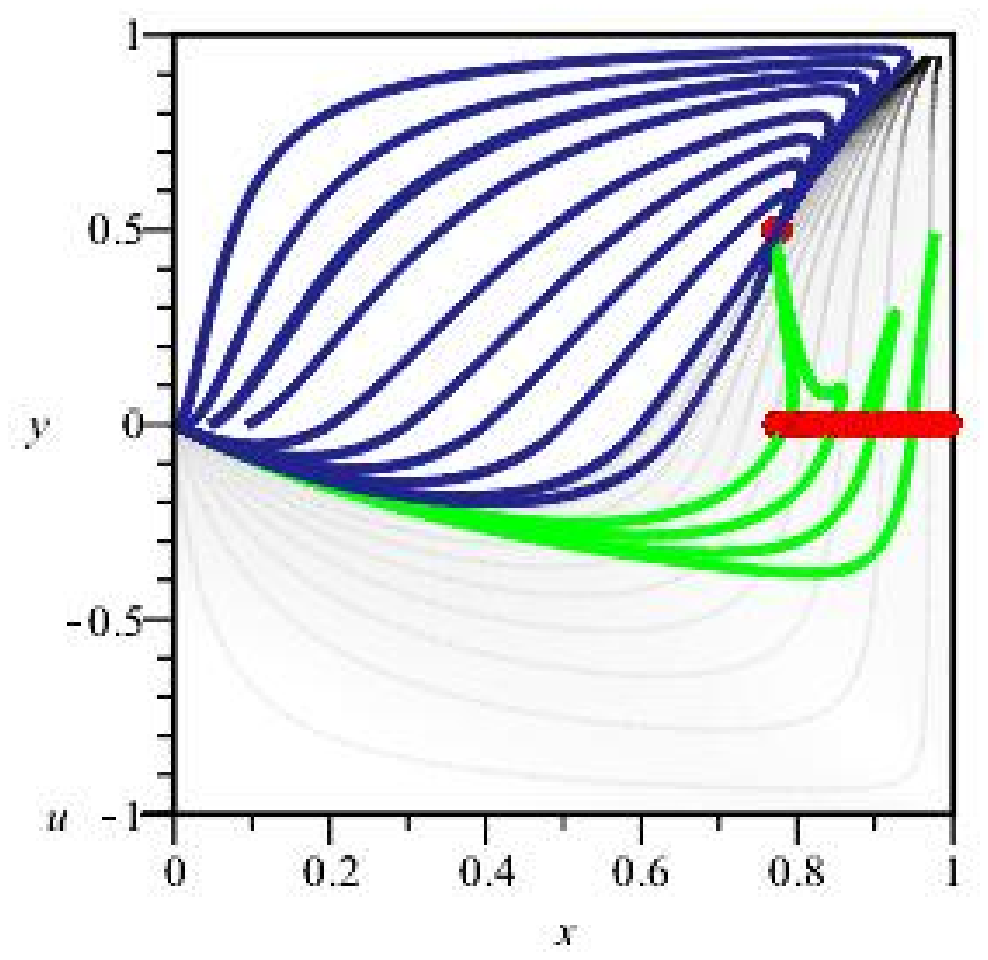}
\vspace{1.2cm}\caption{Phase portrait of the dynamical system \eqref{mu4-ode} for $\epsilon=+1$ and the following values of the free parameters: $\alpha=1$, $\beta=1$. Two sets of orbits that are generated by chosen initial data are shown in different colors. The boundary of the physically meaningful phase space $F^2=0$, where $F$ is defined in \eqref{bi-frg} -- surface with contours -- is included. In the left and center figures the orientation is given by $[\theta,\phi,\psi]=[-130^o,70^o,10^o]$ and $[\theta,\phi,\psi]=[130^o,70^o,-10^o]$, respectively, where $\theta$, $\phi$ and $\psi$ are the Euler angles. Meanwhile, in the right figure the $x,y$-projection is shown: $[\theta,\phi,\psi]=[-90^o,0^o,0^o]$. The isolated point represents the de Sitter attractor $P_\text{dS}$, while the dotted curved segment represents the de Sitter critical manifold ${\cal P}_\text{dS}$.}\label{fig2}\end{figure*}

%----------------------------------------

%==================================

\subsection{The 3D phase portrait}

The 3D phase space where to look for phenomenologically viable behavior of the dynamical system \eqref{mu4-ode} is defined in the following way:

\bea \Psi_\text{3D}=\left\{(x,y,u):0\leq x\leq1,-1\leq y\leq 1,0\leq u\leq 1,u\leq u_*^\pm\right\},\label{3d-ps}\eea where

\bea u_*^\pm=\frac{\left[x(1-x)-\frac{2\beta}{3\alpha^2}x^2\right](1\mp y_\pm)+(1-x)y_\pm}{\left[(1-\epsilon)x(1-x)-\frac{2\beta}{3\alpha^2}x^2\right](1\mp y_\pm)+(1-x)y_\pm},\nonumber\eea and, as before, $y=y^-\cup y^+$.

The 3D phase portrait of the dynamical system \eqref{mu4-ode} is shown in FIG. \ref{fig2} for $\epsilon=1$ and the following values of the free parameters: $\alpha=1$, $\beta=1$. Two sets of orbits that are generated by chosen initial data sets, are shown in different colors. A surface with contours representing the boundary of the physically meaningful phase space: $F^2=0$, where $F$ is defined in \eqref{bi-frg}, has been included in the figure. In the left and center figures the orientation is given by $[\theta,\phi,\psi]=[-130^o,70^o,10^o]$ and $[\theta,\phi,\psi]=[130^o,70^o,-10^o]$, respectively, where $\theta$, $\phi$ and $\psi$ are the Euler angles. Meanwhile, in the right figures the $x,y$-projection is shown: $[\theta,\phi,\psi]=[-90^o,0^o,0^o]$. The isolated point in each figure represents the de Sitter attractor $P_\text{dS}$, while the dotted curved segment represents the de Sitter critical manifold ${\cal P}_\text{dS}$. 

Notice that the phase space orbits end up either at the isolated de Sitter attractor $P_\text{dS}$, or at local attractor de Sitter points in the manifold ${\cal P}_\text{dS}$ (uppermost points in the manifold). The green orbits end up, precisely, at the uppermost points of ${\cal P}_\text{dS}$, while navy orbits end up at the isolated de Sitter attractor. In any case the state towards the universe is attracted which corresponds to de Sitter expansion $a(t)\propto\exp{H_0 t}$, where $H^2_0$ is given either by \eqref{h2-mds} or by \eqref{p6cond-1}.

%%%%%%%%%%%%%%%%%%%%%%%%%%%%%%%%%%%%%%%%%%%%%%%%%%%%%%%%%%%%%%%%%%%%%%%%%%%%%%%%%%%%%%%%%%%

\section{Avoidance of the cosmological constant issue in the present setup}\label{sect-ccp}

Perhaps the most interesting solution in the phenomenologically viable subspace of the phase space is the de Sitter attractor $P_\text{dS}$ (here we should add the de Sitter points in the critical manifold ${\cal P}_\text{dS}$ which exist only in the case where $\mu^4$ is non-vanishing). The de Sitter attractor solutions are interesting in the present model because these entail that, at late time, the FRW universe described by the theory \eqref{frg-action}, \eqref{bi-frg}, is almost indistinguishable from the $\Lambda$CDM cosmological model. One should not be surprised by this result since, at low curvature, a non-vanishing cosmological constant $\Lambda_\text{eff}=(\lambda^2-2\epsilon\kappa^2\mu^4)/\alpha\lambda^2$, arises in this model. The surprising result is that, at the de Sitter attractor $H^2_0\neq\Lambda_\text{eff}/3$, which challenges our intuition. 

In order to fix ideas, momentarily, we shall choose $\epsilon=+1$. Let us to set $\lambda^2=2\kappa^2\mu^4$ in \eqref{frg-action}. Under this choice our model coincides with the one previously investigated in \cite{comelli}, where an exact cancellation mechanism of the cosmological constant has been applied.\footnote{The asymptotic dynamics of this model was studied in \cite{jcap-2010} for negative Gauss-Bonnet coupling.} Actually, under the above choice the effective cosmological constant at low curvature, $\Lambda_\text{eff}=0$, exactly vanishes. This includes, as a particular case, the flat space ($R={\cal G}=0$), for which:

\bea F(R,{\cal G})=-\lambda^2\sqrt{1-\alpha R-\beta{\cal G}},\;F_R=-\frac{\alpha\lambda^4}{2F}\;\Rightarrow\;F(0,0)=-\lambda^2,\;F_R(0,0)=\frac{\alpha\lambda^2}{2}.\nonumber\eea In this particular case, assuming vacuum background, we have that $G_{\mu\nu}=0$ ($R_{\mu\nu}=0$), while the fourth-order curvature contributions $\Sigma^\text{curv}_{\mu\nu}$ in \eqref{sigma-mn}, amount to: 

\bea \Sigma^\text{curv}_{\mu\nu}(0,0)=-\frac{F(0,0)}{2F_R(0,0)}g_{\mu\nu}=\frac{1}{\alpha}g_{\mu\nu},\nonumber\eea so that the EOM \eqref{feq} for vacuum: 

\bea \Sigma^\text{curv}_{\mu\nu}(0,0)=\frac{\kappa^2\mu^4}{F_R(0,0)}g_{\mu\nu}=\frac{2\kappa^2\mu^4}{\alpha\lambda^2}g_{\mu\nu},\label{vac-feq}\eea become an identity after our choice $2\kappa^2\mu^4=\lambda^2$. Hence, as long as both $F_R(0,0)=\alpha\lambda^2/2>0$ and $F_{RR}(0,0)=\alpha^2\lambda^2/4>0$ are positive quantities, flat space is a stable solution of the EOM of our setup. This is to be contrasted with the $F(R)$ model investigated in \cite{ccp-staro}, where the flat space was a unstable solution of the equations of motion. 

From equation \eqref{feq} it is seen that, for our above choice $\epsilon=+1$, the vacuum energy density is a negative quantity: $\rho_\text{vac}=-\mu^4$. But this is not problematic since, as mentioned above, the effective energy density of vacuum at small curvature vanishes. The fact we want to underline here is that the energy density of vacuum $\rho_\text{vac}$, the effective cosmological constant $\Lambda_\text{eff}$ in the low-curvature regime and the present value of the Hubble rate $H_0$, are unrelated quantities in our setup. Actually, for the de Sitter attractor $P_\text{dS}$ \eqref{mu4-attractor-ds}, that arises in the phase space corresponding to our cosmological model, the constant expansion rate reads:

\bea H_0^2=\frac{\alpha}{4\beta}\left(\sqrt{1+\frac{2\beta}{3\alpha^2}}-1\right).\label{h20-pds}\eea It has nothing to do neither with $\rho_\text{vac}$ nor with $\Lambda_\text{eff}$. Actually, given that $\Lambda_\text{eff}=0$ thanks to our choice ($\lambda^2=2\kappa^2\mu^4$), and that the de Sitter attractor $P_\text{dS}$ arises no matter whether $\mu^4=0$, as in Sec. \ref{sect-bigb-2d}, or $\mu^4\neq 0$, as in Sec. \ref{sect-bigb-3d}, the constant Hubble rate \eqref{h20-pds} is independent of the vacuum energy density $\rho_\text{vac}=-\mu^4$, as well as of the effective (low-curvature) cosmological constant $\Lambda_\text{eff}=0$. This non-trivial fact is at the core of the avoidance of the cosmological constant issue in the present setup.

It is apparent from above that the theory \eqref{frg-action} with $F(R,{\cal G})$ given by \eqref{bi-frg} and $\lambda^2=2\kappa^2\mu^4$ ($\epsilon=+1$), satisfies the necessary and sufficient conditions discussed in the introduction (Sec. \ref{sect-intro}), as well as the additional reasonable requirements, that are to be satisfied in order to have a phenomenologically satisfactory theory of gravity where the cosmological constant problem does not arise. There are, however, certain observational constraints that should be satisfied as well. Take, for instance, the bond imposed on the mass of the scalar perturbation \cite{newton-law} around flat space \eqref{newton-law-bond}:

\bea m^2_0=\frac{F_R(0,0)}{3F_{RR}(0,0)}=\frac{2}{3\alpha}>2.5\times 10^{-59}\;M^2_\text{Pl}\Rightarrow\alpha<10^{60}\;M^{-2}_\text{Pl}.\label{vac-newton-bond}\eea 

Let us consider two limiting situations.

%===============================================================================

\subsection{Vanishing Gauss-Bonnet contribution}

%==============================================================================

Assume, first, that the dimensionless quantity $\beta/\alpha^2\ll 1$ is very small (formal limit $\beta\rightarrow 0$). This means that the Gauss-Bonnet contribution is negligible and we are left with an $F(R)$ theory. In this case from \eqref{h20-pds} it follows that:

\bea H^2_0\approx\frac{1}{12\alpha}>10^{-60}\;M^2_\text{Pl},\nonumber\eea which means that the observational constraint $H^2_0\sim 10^{-120}\,M^2_\text{Pl}$ on the present value of the Hubble rate, can not be satisfied unless one gives up the requirement \eqref{vac-newton-bond}. As a matter of fact this requirement can be smoothed out or even avoided due to the chameleon effect \cite{cham-khoury, cham-quiros}: The effective mass of the scalar degree of freedom may be a function of the local background curvature or, equivalently, of the energy density of the local environment, so that it can be large at Solar System and terrestrial curvatures and densities and small at cosmological scales. In other words: it may be short ranged in the Solar System and become long ranged at cosmological densities, thus affecting the cosmological dynamics. It has been shown that for metric $F(R)$ theories the chameleon effect plays an important role \cite{sotiriou-faraoni, cham-cembranos, cham-navarro, cham-faulkner}. In this $F(R)$ limit of the present formalism there are not Laplacian or gradient instabilities.

%==================================================================================

\subsection{Dominant Gauss-Bonnet contribution}

%=================================================================================

The other limiting situation $\beta/\alpha^2\gg 1$ or, formally, $\alpha\rightarrow 0$, is when the Gauss-Bonnet term dominates the late times cosmological dynamics. From \eqref{h20-pds} it follows that:

\bea H^2_0\approx\frac{1}{2\sqrt{6\beta}},\nonumber\eea so that the experimental bond \eqref{vac-newton-bond} is avoided in this case. The price to pay for evading the constraint \eqref{vac-newton-bond} is the strong Gauss-Bonnet coupling required: $\beta\sim 10^{240}\,M^{-4}_\text{Pl}$. This sets the scale of smallness of the Gauss-Bonnet term $|{\cal G}|$ much below $1/\beta\sim 10^{-240}\,M^4_\text{Pl}$, i. e., the scale of small curvature $\sqrt{|{\cal G}|}$ is far below the present value of the curvature of the Universe $\sim 10^{-120}\,M^2_\text{Pl}$. This is why, in this limit, the present cosmological dynamics is dictated by the Gauss-Bonnet interaction. At much smaller curvature scales, $|{\cal G}|\ll 1/\beta$, the Gauss-Bonnet term amounts to a total derivative and may be safely removed, so that the coupling $\beta$ does not play any role.

In both limiting situations the resulting physical picture is one in which the energy density of vacuum is of the order of the Plack mass to the 4th power, with vanishing effective (low-curvature) cosmological constant because flat space is a stable solution of the vacuum EOM, and a FRW de Sitter expansion with the required (present day) Hubble rate $H_0$, is the global future attractor.

%-----------------------------------

\begin{figure*}[t!]
\includegraphics[width=7cm]{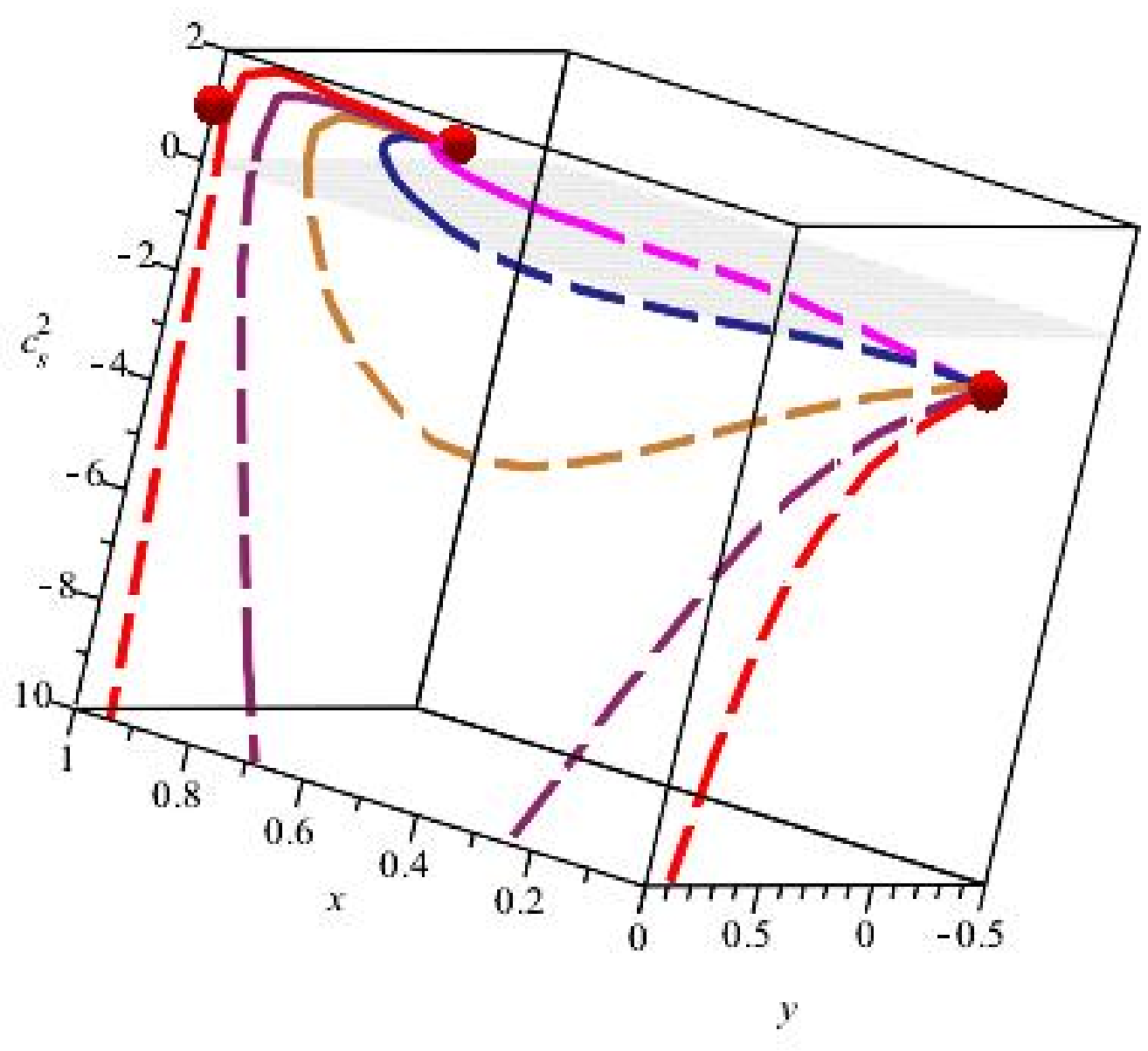}
\includegraphics[width=7cm]{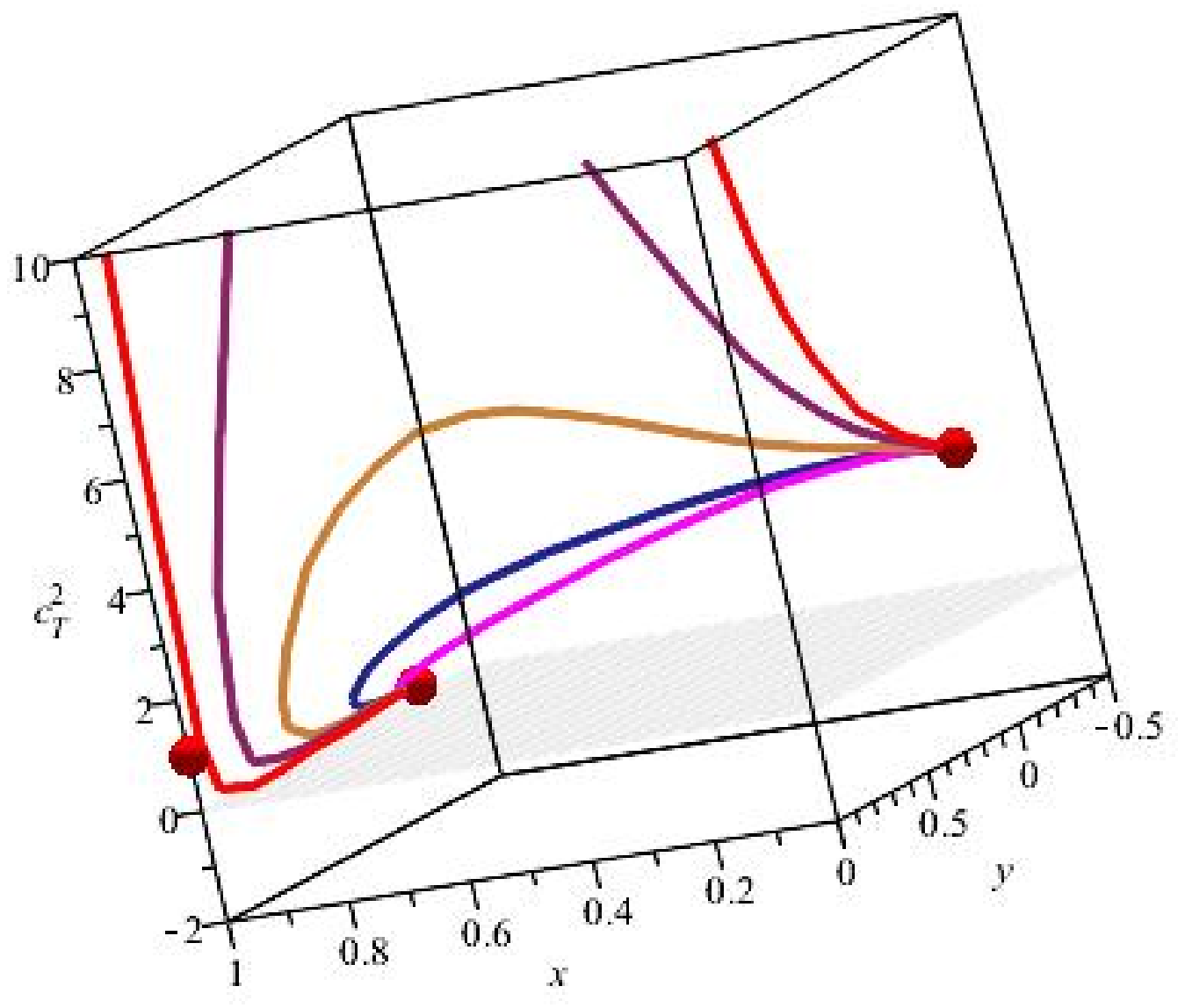}
\vspace{1.2cm}\caption{Plots of the squared speed of propagation of scalar (left) and tensor (right) perturbations vs the coordinates of the phase plane -- $c^2_s=c^2_s(x,y)$ and $c^2_T=c^2_T(x,y)$, respectively -- for the orbits in the center panel of FIG. \ref{fig1} ($\mu^4=0$, $\alpha=\beta=1$). The dashed parts of the curves mean that the squared speed of propagation is negative, signaling the occurrence of a Laplacian instability.}\label{fig3}\end{figure*}

%----------------------------------------

%%%%%%%%%%%%%%%%%%%%%%%%%%%%%%%%%%%%%%%%

\section{Discussion}\label{sect-discuss}

%%%%%%%%%%%%%%%%%%%%%%%%%%%%%%%%%%%%%%%%

The absence of ghosts and instabilities such as: ghosts due to anisotropy of space or to linear perturbations around spherically symmetric static background \cite{spheric-perts}, tachyonic, Dolgov-Kawasaki and graviton instabilities, in the present model is a consequence of the choice of $F(R,{\cal G})$ in \eqref{bi-frg}. This makes of the theory \eqref{frg-action}, with $F(R,{\cal G})$ given by \eqref{bi-frg}, a very attractive possibility for a viable fourth-order theory of gravity. 

Unfortunately, but for the formal limit $\beta\rightarrow 0$, the model is not free of other kinds of problems such as the Laplacian or gradient instability. Their absence would either impose additional requirements on the physical phase space, as well as on the space of parameters of the theory, or require of additional modifications to the original set up. According to \cite{defelice-prd-2010, defelice-jcap-2009}, the squared speed of propagation of the scalar and tensor modes (gravitational waves) in general $F(R,{\cal G})$ theories, are given by:

\bea c^2_s=1+\frac{8\frac{\beta}{\alpha}\dot H}{1+4\frac{\beta}{\alpha}H^2}=1+2(1-x)\frac{\dot H}{H^2},\label{c2s-defelice}\eea and 

\bea c^2_T=\frac{F_R+4\frac{\beta}{\alpha}\ddot F_R}{F_R+4\frac{\beta}{\alpha}H\dot F_R}=1-2(1-x)\frac{\dot H}{H^2},\label{c2t-defelice}\eea respectively. Notice that $c^2_s+c^2_T=2$. For the absence of Laplacian instabilities it is required that both $c^2_s\geq 0$ and $c^2_T\geq 0$, i. e., that

\bea -\frac{1}{2(1-x)}\leq\frac{\dot H}{H^2}\leq\frac{1}{2(1-x)}.\label{no-laplace}\eea Whether $c^2_s>1$ or $c^2_T>1$, the speed of propagation is superluminal. However, this is not a problem as it does not directly violate causality on the cosmological FRW background \cite{defelice-prd-2010}. On the contrary, it is a real problem when either $c^2_s<0$ or $c^2_T<0$ (or both), since a catastrophic gradient (also Laplacian) instability develops. It has been demonstrated in \cite{defelice-prd-2010, defelice-perts} that the Laplacian instability at short wavelength scales is inevitable in general $F(R,{\cal G})$ theories. This includes our present set up. Let us to consider the two limiting situations studied in section \ref{sect-ccp}: i) the $F(R)$ limit where $\beta\rightarrow 0$ and ii) the Gauss-Bonnet dominance $\alpha\rightarrow 0$. In the former case we have that 

\bea c^2_s=c^2_T=1,\nonumber\eea while in the latter limiting situation:

\bea c^2_s=1+2\frac{\dot H}{H^2},\;c^2_T=1-2\frac{\dot H}{H^2},\nonumber\eea confirming that in the $\beta\rightarrow 0$ limit there are not Laplacian instabilities (the speed of propagation of both scalar and tensor modes coincides with the speed of light), while in the $\alpha\rightarrow 0$ limit these instabilities are inevitable.

As an illustration of the inevitable appearance of Laplacian or gradient instabilities in our set up, in FIG. \ref{fig3} the plots of $c^2_s=c^2_s(x,y)$ (left) and of $c^2_T=c^2_T(x,y)$ (right) are shown for several orbits of the dynamical system \eqref{ode}, corresponding to the simplified case when $\mu^4=0$ (see sub-section \ref{subsect-0vac}), and to the choice of free parameters: $\alpha=\beta=1$ (middle panel of FIG. \ref{fig1}). It is seen that, although $c^2_T$ is always a positive quantity, the squared speed of the scalar perturbations $c^2_s$ is negative along parts of the orbits, signaling that Laplacian instabilities are inevitable. This has been associated with matter instabilities that arise at small scales and large redshifts in $F(R,{\cal G})$ theories \cite{defelice-prd-2010, defelice-perts}.

One possibility to deal with these unwanted instabilities can be based on the method developed in \cite{nojiri-ghfree-grav}, where the authors proposed a procedure to eliminate ghosts and to obtain viable $F(R,{\cal G})$ models. The method is based on the introduction of an auxiliary scalar field into the $F(R,{\cal G})$ action. Then, in order to make the scalar mode not a ghost, a canonical kinetic term may be introduced in the action. After this, it is possible to obtain second-order equations of motion and to impose suitable initial conditions determining a regular (unique) ghost-free evolution. Another possibility can be to introduce non-minimal coupling to a scalar field $\phi$ of the following form (compare with \eqref{pwl-frg}):

\bea F(\phi,R,{\cal G})=-\lambda^2\left(1-\alpha\phi^{-1}R-K(\phi)-\beta{\cal G}\right)^\nu,\nonumber\eea where $K(\phi)$ is a kinetic term for the scalar field. The coupled metric and scalar perturbations might conspire to counteract the effects of the destabilizing Gauss-Bonnet contributions. We do not expect that the present setup may account for a realistic description of our Universe, instead, it may be viewed as a toy model showing that the cosmological constant issue, if not solved, at least may be evaded.

%------------------------------------------------------------------------------------------------

Although our model complies the requirements mentioned in the introduction, nevertheless, there should be other models of modified gravity that may fulfill these requirements as well. The proposal investigated in \cite{ccp-staro} seems to be an example of that. In that reference the author studies $F(R)$ models given by:

\bea F(R)=R+\lambda R_0\left[\left(1+\frac{R^2}{R_0^2}\right)^{-n}-1\right],\label{staro-fr}\eea where $n$, $\lambda>0$ and $R_0$ of the order of the presently observed effective cosmological constant, are the free parameters. Then, $F(0)=0$ (the cosmological constant ``disappears'' in flat space) and $R_{\mu\nu}=0$ is always a solution of the EOM in the absence of matter. For $R\gg R_0$, $F(R)=R-2\Lambda_*$, where $\Lambda_*=\lambda R_0/2$. The model has de Sitter solutions with $R=x_1R_0$, where $x_1$ is the maximal root of a given algebraic equation (equation (6) of \cite{ccp-staro}). This model is free of the Dolgov-Kawasaki instability, unfortunately, given that $F_{RR}<0$, flat space is unstable. For $n\geq 2$ the model passes laboratory and Solar System tests of gravity. In order to check that this proposal meets the necessary and sufficient conditions stated in the introduction it is mandatory to consider an action of the kind \eqref{frg-action} with $F(R)\rightarrow F(R,{\cal G})$. I. e., from the start a non-vanishing vacuum energy density $\propto\mu^4$ should be considered. Besides, it should be demonstrated that the de Sitter solution with $R=x_1R_0$ is a future attractor in the phase space of the model. To our knowledge these items have not been investigated yet.

%%%%%%%%%%%%%%%%%%%%%%%%%%%%%%%%%%%%%%%%

\section{Conclusion}\label{sect-conclu}

%%%%%%%%%%%%%%%%%%%%%%%%%%%%%%%%%%%%%%%%

In this paper we have explored a class of Born-Infeld inspired $F(R,{\cal G})$ models of modified gravity of type $F(R,{\cal G})\propto\sqrt{{\cal L}_\text{Lovelock}}$ (see equation \eqref{bi-frg}):

\bea F(R,{\cal G})=-\lambda^2\sqrt{1-\alpha R-\beta{\cal G}}.\nonumber\eea This two-parametric class of theory is free of most of the unwanted ghosts and instabilities, but for the Laplacian instability which is inevitable in $F(R,{\cal G})$ models in general. This calls for further modifications of \eqref{bi-frg} through, for instance, the procedure proposed in \cite{nojiri-ghfree-grav}. 

Models of the kind we have investigated here are very interesting alternatives to GR, not only because the inclusion of higher-order curvature operators is dictated by renormalization \cite{stelle, hindawi} and by the formulation of GR as an effective theory \cite{donoghue-1, donoghue-2}, but because the CCP may be avoided in these models. It should be mentioned that the full asymptotic dynamics of the present model has been investigated in \cite{jcap-2010}. In that reference, however, only the case with negative Gauss-Bonnet coupling ($\beta<0$) was considered. Although this latter case was not investigated here, the results of our present study also apply to this case. 

It is necessary to mention, also, that the formal limit when in model \eqref{bi-frg} $\beta\rightarrow 0$, i. e., if neglect the Gauss-Bonnet term, will not satisfy the observational constraints coming from cosmology, in particular that the present value of the Hubble rate $H_0\sim 10^{-60}\;M_\text{Pl}$, unless the chameleon effect plays an important part in the origin of the effective mass of the propagating scalar degree of freedom. The advantage of this limiting case is that Laplacian instability does not arise. Alternatively, when the Gauss-Bonnet interaction plays a significant role in the late time dynamics, the observational constraints of cosmological origin are met but the model should be improved in order to avoid the arising of gradient instability. 

Although the class of theory \eqref{bi-frg} meets the necessary and sufficient conditions to avoid the cosmological constant issue, there remain issues with the occurrence of Laplacian instabilities. Hence, in the last instance one may think of the present setup as a class of toy models that serve to show that, an alternative to explain the huge discrepancy between the theoretically predicted and the observed values of the cosmological constant, does exist. It consists just in evading the problem.

%%%%%%%%%%%%%%%%%%%%%%%%%%%%%%%%%%%%%%%%%%%%%%

\section*{Acknowledgments}

%%%%%%%%%%%%%%%%%%%%%%%%%%%%%%%%%%%%%%%%%%%%%%

The author thanks FORDECYT-PRONACES-CONACYT for support of the present research under grant CF-MG-2558591.

%%%%%%%%%%%%%%%%%%%%%%%%%%%%

%%%%%%%%%%%%%%%%%%%
%%%%%%%%%%%%%%%%%%

\end{document}